\documentclass[twocolumn,superscriptaddress,aps,prl,longbibliography,showpacs,preprintnumbers,amsmath,amssymb,floatfix]{revtex4-1}
\usepackage{titlesec}
\usepackage{amsmath}
\usepackage{graphicx}
\usepackage[ansinew]{inputenc}
\usepackage{array}
\usepackage{color}
\usepackage{amsxtra}
\usepackage{amstext}
\usepackage{amssymb}
\usepackage{latexsym}
\usepackage{gensymb}
\usepackage{dsfont}
\usepackage{braket}
\usepackage[caption=false]{subfig}
\usepackage[makeroom]{cancel}

\usepackage{dcolumn}
\usepackage{bm}
\usepackage{bbm}
\usepackage{braket}
\usepackage{mathrsfs}
\usepackage{amstext}
\usepackage{euscript}
\usepackage{txfonts}

\usepackage[unicode=true,
 bookmarks=false,
 breaklinks=false,pdfborder={0 0 1},backref=false,colorlinks=true,citecolor=blue]
 {hyperref}

\newcommand{\ms}[1]{\mbox{\scriptsize #1}}

\makeatletter
\@ifundefined{textcolor}{}
{%
 \definecolor{BLACK}{gray}{0}
 \definecolor{WHITE}{gray}{1}
 \definecolor{RED}{rgb}{1,0,0}
 \definecolor{GREEN}{rgb}{0,1,0}
 \definecolor{BLUE}{rgb}{0,0,1}
 \definecolor{CYAN}{cmyk}{1,0,0,0}
 \definecolor{MAGENTA}{cmyk}{0,1,0,0}
 \definecolor{YELLOW}{cmyk}{0,0,1,0}
}

\makeatletter
\renewcommand*\env@matrix[1][*\c@MaxMatrixCols c]{%
  \hskip -\arraycolsep
  \let\@ifnextchar\new@ifnextchar
  \array{#1}}
\makeatother

\newcommand{\eref}[1]{Eq.\,\eqref{#1}}
\newcommand{\fref}[1]{Fig.\,\ref{#1}}

\newcommand{\cref}[1]{Ref.\,\cite{#1}}

\newcommand{\mc}[1]{\mathcal{#1}}

\makeatother

\begin{document}

\title{Distributed Quantum Metrology and the Entangling Power of Linear Networks}

\author{Wenchao Ge}
\affiliation{United States Army Research Laboratory, Adelphi, Maryland 20783, USA}
\affiliation{The Institute for Research in Electronics and Applied Physics (IREAP), College Park, Maryland 20740, USA}

\author{Kurt Jacobs}
\affiliation{United States Army Research Laboratory, Adelphi, Maryland 20783, USA}
\affiliation{Department of Physics, University of Massachusetts at Boston, Boston, Massachusetts 02125, USA}
\affiliation{Hearne Institute for Theoretical Physics, Louisiana State University, Baton Rouge, Louisiana 70803, USA}

\author{Zachary Eldredge}
\affiliation{Joint Quantum Institute, NIST/University of Maryland, College Park, Maryland 20742, USA}
\affiliation{Joint Center for Quantum Information and Computer Science, NIST/University of Maryland, College Park, Maryland 20742, USA}

\author{Alexey V. Gorshkov}
\affiliation{Joint Quantum Institute, NIST/University of Maryland, College Park, Maryland 20742, USA}
\affiliation{Joint Center for Quantum Information and Computer Science, NIST/University of Maryland, College Park, Maryland 20742, USA}

\author{Michael Foss-Feig}
\affiliation{United States Army Research Laboratory, Adelphi, Maryland 20783, USA}
\affiliation{Joint Quantum Institute, NIST/University of Maryland, College Park, Maryland 20742, USA}
\affiliation{Joint Center for Quantum Information and Computer Science, NIST/University of Maryland, College Park, Maryland 20742, USA}

%\date{\today}
\begin{abstract}
We derive a bound on the ability of a linear optical network to estimate a linear combination of independent phase shifts by using an arbitrary non-classical but \emph{unentangled} input state, thereby elucidating the quantum resources required to obtain the Heisenberg limit with a multi-port interferometer. Our bound reveals that while linear networks can generate highly entangled states, they cannot effectively combine quantum resources that are well distributed across multiple modes for the purposes of metrology: in this sense linear networks endowed with well-distributed quantum resources behave classically.  Conversely, our bound shows that linear networks can achieve the Heisenberg limit for distributed metrology when the input photons are hoarded in a small number of input modes, and we present an explicit scheme for doing so. Our results also have implications for measures of non-classicality.  
\end{abstract}

\maketitle 

By taking advantage of the quantum mechanical properties of micro- and mesoscopic systems, it is possible to increase the sensitivity of precision sensors beyond classical limitations~\cite{caves1981quantum, Holland:1993aa, giovannetti2004quantum, giovannetti2006quantum, campos2003optical,pezze2008mach, cooper2010entanglement, birrittella2012multiphoton, gagatsos2016gaussian}. Recently there has been increasing interest in understanding how such quantum metrological techniques can be used to enhance measurements that are spatially distributed. In this setting, the quantity of interest is often a linear combination of the results of a number of simultaneous measurements at different locations~\cite{Zach:16,proctor2017networked}. Examples of this problem, referred to as \textit{distributed metrology}, are the inference of a field gradient or properties of the spatial fluctuations of a field. To obtain a quantum advantage in distributed metrology, the spatially distributed sensors must share entanglement and thus possess a joint quantum state distributed to them by a quantum network. We note that distributed quantum metrology reduces to the usual (local) quantum metrology when each location measures the same quantity. Thus the results presented here for distributed metrology automatically apply to the usual ``local'' metrology in which each sensor measures the same phase shift. 

\begin{figure}[t]
\leavevmode\includegraphics[width = 1.0 \columnwidth]{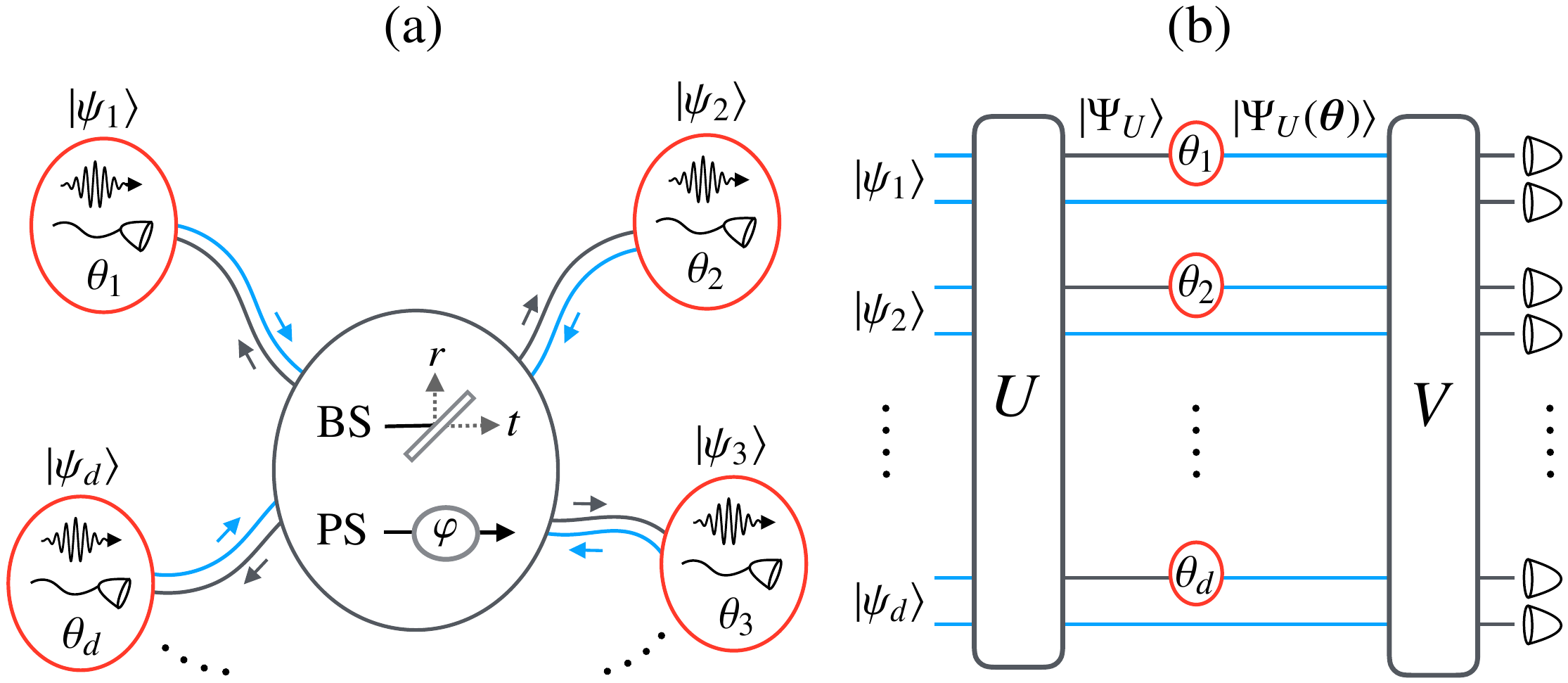}
\caption{(a) Spatial layout of a network of sensors designed to measure a linear combination of $d$ spatially-distributed phase shifts. Each node is equipped with a measurement device and a locally prepared nonclassical state, which can be sent to a central linear network prior to probing the local phase shifts, and once again before making local measurements. (b) Formal representation of the situation described in (a):  To make a fair comparison to the independent metrology of each phase, for which each node would require an additional reference mode in order to form an interferometer, each node is endowed with two input modes.  These $2d$ modes are fed into a unitary $U$ and the phases are then probed, after which we allow further linear-optical processing ($V$) followed by local measurements.} 
\label{fig:layout} 
\end{figure} 

It is well known that linear-optical networks consisting purely of beam splitters and phase shifters can transform nonclassical but unentangled states into highly mode-entangled states~\cite{Kim:02,Wang:02,Wolf:03,Asboth:05,Vogel:14, ge2015conservation, Killoran:2016aa}. Given that linear networks are also quite easily implemented experimentally, much attention has been paid to applications of the entanglement they can generate, for example towards demonstrations of quantum supremacy via boson sampling \cite{Aaronson_2013,Tillman_2013,Broome794,Spring798,Crespi2013}. However, it is also clear that the entangling power of linear optical networks is limited; given as input a particular quantum state, they cannot produce \emph{arbitrary} quantum states at the output \cite{kok2000limitations, Kok:2007aa}. It is therefore natural to ask whether---for a particular application requiring a particular type of entanglement---linear optical networks can or can not provide a quantum advantage.  In this manuscript, we investigate whether or not linear networks can take unentangled inputs and create entangled states capable of providing a quantum advantage in the task of distributed metrology.

While our results have implications for distributed metrology in a rather general context, it is useful to consider the following concrete scenario depicted in \fref{fig:layout}(a). Imagine that there are $d$ observers at different locations in space (which we will call ``nodes''), and that each has the ability to generate a nonclassical state $|\psi_j\rangle$ with at most $n$ photons and use this state to measure a local phase shift $\theta_j$ ($j=1,\dots,d$). Furthermore suppose that rather than ascertaining all of the individual phases, we wish to estimate a particular linear combination $ q  = \sum_j w_j \theta_j$ for some set of weights $\bm{w} = (w_1,\ldots, w_d)$ (this linear combination could be, e.g., an average, or the overlap of the phases $\theta_j$ with some spatial mode). To make a comparison with the metrology of a single quantity straightforward, we normalize the weights such that $\max_j|w_j|=1/d$.  With this normalization, choosing the $w_j$ all equal to each other recovers the spatial average of the fields, $ q =\frac{1}{d}\sum_{j=1}^{d}\theta_j$. If the observers are each allocated $n$ photons but no network exists for sharing their quantum states, the best strategy for estimating $ q $ is for each of them to make the best possible estimate of $\theta_j$, and then compute $ q $ by sharing their results via classical communication. We assume that each individual node has access to a reference mode with no phase shift, and that the measurement of $\theta_j$ can be made at the Heisenberg limit, $\Delta\theta_j\sim1/n$.  If all of the phases $\theta_j$ (or at least a number of them that scales with $d$) contribute meaningfully to $q$, a notion captured formally by the requirement $|\bm{w}|^2 \sim 1/d$, then we will say that the weights are \emph{well-distributed}.  Standard error analysis shows that in the case of well-distributed weights, the ``classical'' \footnote{Note that by classical we imply only a lack of correlations between the nodes, while the estimation procedure employed at each node is \emph{not} assumed to be classical.  The natural definition of ``classical'' might be augmented to allow classical distribution of the initial resources, but this will not improve the scaling of $\Delta q _{\rm cl}$ for well-distributed $\bm{w}$.} estimation strategy just described produces an error in the estimate of $ q $ scaling as $\Delta q_{\rm cl}\sim1/(n\sqrt{d})$.  On the other hand, a fully quantum protocol using an optimal entangled state of all the modes that contains a total of $n d$ photons can achieve a scaling of $\Delta q_{\ms{q}} = 1/(nd)$~\cite{proctor2017networked}, thus providing a collective quantum enhancement proportional to $1/\sqrt{d}$. In the context of the above scenario, our central question is as follows: If the observers are allowed to share their initially unentangled states through a central linear network, can they beat the aforementioned classical limit in measuring $q$ for well-distributed weights, in the sense of enhanced scaling with respect to $d$? If so, we wish to know what kinds of unentangled input states we must send into a linear network to obtain output states that can be used to obtain a measurement accuracy scaling at the Heisenberg limit.  

Our conclusion is that the ability of a linear network to generate entanglement that is useful for quantum metrology is strongly bounded by specific properties of the input states.
%This bound applies to an arbitrary set of input states, and in doing so reveals a fundamental limitation of the entangling power of linear transformations.
Our results are most simply stated by first considering the case in which the $j^{\rm th}$ node possesses a state with exactly $n_j$ photons. In this case, and denoting by $\bm{n}$ the vector of input photon numbers $n_j$, we show that the metrological accuracy that can be achieved for measuring $q$ satisfies
\begin{align}
\label{eq:bound}
\Delta q\geq d|\bm{w}|^2/\left(2|\bm{n}|\right).
\end{align}
Two important conclusions can be drawn from \eref{eq:bound}.  First, if in addition to the weights being well distributed ($|\bm{w}|^2\sim 1/d$) the photons are also well distributed, for example if each node has exactly $n$ photons, then we have $\Delta q\gtrsim 1/(n\sqrt{d})$. Considering this scaling (i.e., ignoring any prefactors), and comparing to the classical scheme above, we conclude that linear networks cannot generate useful entanglement for the purposes of combining metrological resources across modes, thereby
%This result is surprising in view of the fact that linear networks have been shown to generate useful entanglement for the purposes of computation \comment{What specifically do we mean here? I'm not sure what the computational power of linear optics is in the absence of measurement and feedback, though presumably it surpasses classical computation for some tasks.}, and reveals that there is a non-trivial relationship between the classes of entanglement that are nonclassical as measured by these two functions. It also shows
establishing a strong operational sense in which linear networks are classical.  Second, if the same total number ($N=nd$) of photons are distributed across a finite number of input ports, for example just two, the bound reduces to $\Delta q\gtrsim1/(nd)$.  Thus, if the bound is tight (at least in the sense of scaling), then Heisenberg scaling for $\Delta q$ can be achieved so long as the total number of photons employed is placed in a small number of modes (``small'' meaning not scaling $d$).  By constructing an explicit measurement scheme we confirm that the bound, and thus Heisenberg scaling, can be saturated when the total number of photons is divided evenly between just two input modes. We also note that our bound motivates certain notions of non-classicality, and we discuss this further below. 

\emph{Multi-parameter quantum Cram\'er-Rao bound.}---A formal analysis of the scenario depicted in \fref{fig:layout}(a) is summarized in \fref{fig:layout}(b).  Note that we have now explicitly introduced one reference mode per phase $\theta_j$, with index $d+j$.  We assume that the input state is a product state $\ket{\Psi}=\ket{\psi_1}\dots\ket{\psi_d}$ and that each $\ket{\psi_j}$ is itself a product state between the two input modes associated with phase $\theta_j$, $\ket{\psi_{j}}=\ket{\varphi_{j}}\ket{\varphi_{d+j}}$, so that $\ket{\Psi}=\otimes_{j=1}^{2d}\ket{\varphi_j}$.  The state that the linear network produces is then denoted by $|\Psi_{U}\rangle=U\ket{\Psi}$.  Defining $\hat{H}=\sum_{j}\theta_j\hat{n}_j$, interrogation of the phases $\theta_j$ then maps $\ket{\Psi_U}\rightarrow \exp(-i\hat{H})\ket{\Psi_{U}}\equiv \ket{\Psi_U(\bm{\theta})}$.  We aim to infer the quantity $ q =\sum_{j}w_j\theta_j$ by making local measurements on $\ket{\Psi_U(\bm{\theta})}$, preceded (if desired) by an additional linear-optical unitary $V$.  The primary tool in our analysis is the multi-parameter quantum Cram\'er-Rao bound, which states that a set of unbiased estimators $\Theta_j$ for the parameters $\theta_j$ satisfy \footnote{This inequality is short-hand for the asymptotic per-trial covariance.  Strictly speaking, for $\nu$ trials of an estimation procedure we have ${\rm Cov}(\bm{\Theta})\geq \left(\nu\mathcal{F}\right)^{-1}$, which is saturable in the large $\nu$ limit.} 
\begin{align}
\label{eq:qcrb}
{\rm Cov}(\bm{\Theta})\geq \mathcal{F}^{-1}.
\end{align}
Here the covariance matrix is defined by its matrix elements as ${\rm Cov}(\bm{\Theta})_{jk}\equiv E[(\Theta_j-\theta_j)(\Theta_k-\theta_k)]$, where $E[X]$ is the expected value of the quantity $X$, and in this context the quantum Fisher information matrix $\mathcal{F}$ is defined by its matrix elements as
\begin{align}
\label{eq:qfi}
\mathcal{F}_{jk}\equiv 4\big(\bra{\Psi_U}\hat{n}_j\hat{n}_k\ket{\Psi_U}  - \bra{\Psi_U}\hat{n}_j\ket{\Psi_U}\bra{\Psi_U}\hat{n}_k\ket{\Psi_U}\big).
\end{align}
Using $Q=\sum_{j}w_j\Theta_j$ as an unbiased estimator of $ q $, the uncertainty $\Delta^2q\equiv E[(Q-q)^2]=\sum_{j,k}w_j{\rm Cov}(\bm{\Theta})_{jk}w_k$, is bounded by \eref{eq:qcrb} as \cite{Zach:16,Paris_2009,proctor2017networked},
\begin{align}
\label{eq:boundtheta}
\Delta^2q\geq\sum_{j,k}w_j\mathcal{F}^{-1}_{jk}w_k.
\end{align}
Note that $\mathcal{F}$ is a real, symmetric, positive \emph{semi}-definite matrix, and need not be invertible in general.  However, in the case that $\mathcal{F}$ is not invertible, the estimation procedure will only succeed if $\bm{w}$ has vanishing projection onto the  kernel of $\mathcal{F}$.  In that case, $\mathcal{F}^{-1}$ should be interpreted as the inverse of $\mathcal{F}$ \emph{after} projection onto the subspace spanned by eigenvectors with non-zero eigenvalues \cite{proctor2017networked}. Thus we hereafter assume that $\mathcal{F}$ has been projected in this manner, and therefore is positive definite (as opposed to positive semi-definite) and invertible.  The bound in \eref{eq:boundtheta} is tight, in the sense that it is guaranteed to be saturable for some choice of a measurement protocol.  However, to obtain our result it will be useful to further bound the r.h.s of \eref{eq:boundtheta} by something more easily computable for a general unitary $U$.  To this end, we use the Cauchy-Schwarz inequality to write
\begin{align}
\sum_{j,k} w_j \mathcal{F}^{-1}_{jk}w_k  \sum_{l,m}w_l\mathcal{F}_{lm}w_m\geq |\bm{w}|^4.
\end{align}
Defining $\mathcal{F}_{\bm{w}}\equiv \sum_{j,k}w_j\mathcal{F}_{jk}w_k$, we then obtain the bound
\begin{align}
\label{eq:q_bound}
\Delta^2 q\geq\frac{|\bm{w}|^4}{\mathcal{F}_{\bm w}}.
\end{align}
Note that if we have at most $n$ photons per mode after applying the unitary $U$, we can write $\mathcal{F}_{\bm{w}}\leq n^2\sum_{j,k}|w_j||w_k|\leq n^2$, which for well distributed weights ($|\bm{w}|^2\sim1/d$) gives $\Delta^2q\gtrsim1/(nd)^2$.  Since $nd$ is the maximum total number of photons, this coincides with the usual Heisenberg limit for measuring a single phase shift.  However, whether or not $\mathcal{F}_{\bm{w}} \propto  n^2$ can actually be achieved depends on the details of the states $|\Psi_U\rangle$ that a linear network can produce.  As we demonstrate below, this in turn depends on the types of nonclassical states $\ket{\psi_j}$ that we have access to at the inputs.

\emph{Fisher information in linear-optical networks}.---The next step in bounding $\Delta q$ is to obtain a bound on $\mathcal{F}_{\bm w}$ in terms of the $2d$ input states.  If we denote the annihilation operators for the $2d$ input (output) modes by $a_j$ ($b_j$), then the action of the network is described by the relation $b_j^\dagger = \sum_k U_{jk} a_k^\dagger$, where $U_{jk}$ are the elements of a $2d\times2d$ unitary matrix. Since the phase shift $\theta_j$ is applied to the output mode $b_j$, the quantum Fisher information matrix can be computed by inserting the operators $\hat{n}_j=\hat{b}^{\dagger}_j\hat{b}^{\phantom\dagger}_j$ into \eref{eq:qfi}.  Rewriting all operators in terms of the input mode operators and taking the expectation value in the initial product state $\ket{\Psi}=\otimes_{j=1}^{2d}\ket{\varphi_j}$, we obtain
\begin{align}
\label{eq:FQFI_exp}
\mathcal{F}_{jk}=4\!\!\sum_{l,m,r,s}U^{}_{jl}U^{*}_{jm}U^{}_{kr}U^{*}_{ks}\Big(\langle\hat{a}^{\dagger}_l\hat{a}^{}_m\hat{a}^{\dagger}_r\hat{a}^{}_s\rangle-\langle\hat{a}^{\dagger}_l\hat{a}^{}_m\rangle\langle\hat{a}^{\dagger}_r\hat{a}^{}_s\rangle\Big).
\end{align}
Later we will derive a bound on $\mathcal{F}_{\bm{w}}$ that holds for arbitrary separable input states, but it is useful to first consider the simpler situation in which all modes are initialized in Fock states, with the $j^\text{th}$ mode having photon number $n_j$.  In this case $\mathcal{F}_{\bm{w}}$ reduces to
\begin{align}
\label{eq:F_reduce}
\mathcal{F}_{\bm{w}}=4\sum_{j,k}\sum_{r\neq s}n_s(n_r+1)(U^{}_{js}w_jU^{*}_{jr})(U^{}_{kr}w_kU^{*}_{ks}).
\end{align}
The restriction $r\neq s$ can be removed if we replace the equality with ``$\leq$'', because the additional term given by $r=s$ is non-negative.  Defining Hermitian matrices $\mathcal{S}$ and $\mathcal{N}$ such that $\mathcal{S}_{rs}=\sum_{j}U^{}_{js}w_jU^{*}_{jr}$, and $\mathcal{N}_{rs}=\delta_{rs}n_r$, \eref{eq:F_reduce} then takes the following compact form,
\begin{align}
\mathcal{F}_{\bm{w}}\leq 4{\rm Tr}\big[\mc{N}\mc{S}(\mc{N}+1)\mc{S}\big].
\end{align}
Standard trace inequalities \cite{Petz_1994,Mirsky_1975} can now be used to write $\mathcal{F}_{\bm{w}}\leq\sum_j{\rm eigs}(\mc{N})_j({\rm eigs}(\mc{N})_j+1){\rm eigs}(\mc{S})_j^2$, where ${\rm eigs}(M)$ is a list of the eigenvalues of the matrix $M$, sorted by absolute value.  The eigenvalues of $\mc{N}$ are clearly $n_j$, and because $\mc{S}$ is a unitary transformation of the matrix ${\rm diag}(\bm{w})$ it has eigenvalues $w_j$.  Therefore, remembering that $\max_j |w_j|=1/d$, we have
\begin{align}
\mathcal{F}_{\bm{w}}\leq 4|\bm{n}|^2/d^2.
\end{align} 
Plugging this bound on the quantum Fisher information into \eref{eq:q_bound} and taking the square root of both sides, we obtain \eref{eq:bound}.  As mentioned earlier, the most important consequences of this bound are that: (i) For well-distributed resources---that is when $n_j\leq n$ such that $n$ that does not scale with $d$---we have $\Delta q\gtrsim 1/(n\sqrt{d})$; thus, the bound proves that a linear optical network endowed with well-distributed resources cannot improve upon the scaling of a classical scheme in which the estimates of all $d$ phases are made independently.  (ii) If the same maximum number ($N=nd$) of photons are all placed in a finite number of input ports, and assuming the bound can still be saturated, then Heisenberg scaling $\Delta q\sim 1/(nd)$ can be achieved.  With considerably more effort, the bound in Eq. \eqref{eq:bound} can be generalized to the case of arbitrary separable input states.  However, before presenting this generalization, we first give an explicit protocol that saturates the above bound in case (ii), thereby obtaining Heisenberg scaling for distributed metrology in a linear-optical network by ``hoarding'' the resources in a few modes.

\emph{Explicit protocol for hoarded-resource states.}---We now show that Heisenberg scaling $\Delta q\sim 1/(nd)$ can be achieved if the total number of photons, $N=nd$, is split evenly between just two input modes, i.e., $\ket{\Psi}=\ket{N/2}\otimes\ket{N/2}\otimes\ket{0}\otimes\cdots\otimes\ket{0}$. The scheme can be viewed as a generalization of the ``twin Fock state'' proposal in \cref{Holland:1993aa} to the task of distributed quantum metrology.  The basic idea is to find a unitary transformation $U$ that distributes the input twin Fock state between the various modes in a way that explicitly encodes the weights $w_j$.  Upon evolution through the network in \fref{fig:layout}b and subsequent measurement of an operator $\hat{O}$, the sensitivity of estimating $q$ can be obtained through the standard error propagation,
\begin{eqnarray}
\label{eq:error_prop}
\Delta q =\frac{\sqrt{\big\langle\hat{O}^2\big\rangle-\big\langle\hat{O}\big\rangle^2}}{\left| \partial\big\langle\hat{O}\big\rangle/\partial q\right|}.
\end{eqnarray} 
Here, the expectation values are taken with respect to the state $\hat{V}\exp(-i\hat{H})\hat{U}\ket{\Psi}$ at the output of the network. Choosing $V=U^{\dagger}$ \cite{Olson:2016} and measuring the observable $\hat{O}=\ket{\Psi}\bra{\Psi}$, which can be accomplished with photon-number resolving detectors at the $2d$ output ports, we have $\braket{O}=\braket{O^2}=|\bra{\Psi_U}e^{-i\hat{H}}\ket{\Psi_U}|^2$, which for small $q$ becomes
\begin{align}
\big\langle\hat{O}\big\rangle\approx1-\left(\bra{\Psi_U}\hat{H}^2\ket{\Psi_U}-\bra{\Psi_U}\hat{H}\ket{\Psi_U}^2\right).
\end{align}
By choosing $U_{i,1}=U_{i+d,1}=\sqrt{|w_i|/2}$, and $U_{i,2}=-U_{i+d,2}=w_i/\sqrt{2|w_i|}$ for $i=1,2,\cdots, d$,
%$U_{1i}=\sqrt{\eta|w_i|}$ and $U_{2i}=w_i\sqrt{\eta|w_i|}$ for $i=1,2,\cdots, d$, $U_{1,d+1}^{\ast}=-U_{2,d+1}=\sqrt{\eta\sum_{j=1}^dw_j}$, where $\eta=1/\left(1+\left|\sum_{j=1}^dw_j\right|\right)$, 
we can encode each phase with its corresponding weight, obtaining
\begin{eqnarray}
\braket{\hat O}\approx 1-\frac{N(N+2)}{8}q^2.
\end{eqnarray}
Plugging this result into \eref{eq:error_prop} we obtain an uncertainty in estimating $q$ of 
\begin{eqnarray}
\Delta q=\frac{2}{\sqrt{2N(N+2)}},
\end{eqnarray}
which exhibits Heisenberg scaling. Though we have used a $2d$ unitary matrix for estimating $d$ phase shifts with $d$ reference ports, it is in fact possible to simplify the scheme to include only one reference port while maintaining Heisenberg scaling (c.f. Fig. 2 for an example involving two phases).  In implementing the scheme, we require Fock-state inputs \cite{cooper2013experimental}, a linear optical network \cite{Reck:1994aa, marek1997realizable}, and photon-number-resolving detectors \cite{kardynal2008avalanche}.

%Our scheme also works if the Fock states prepared in the two nodes are different, for example with photon numbers $N_1$ and $N_2$. Then the ultimate estimation error is $\Delta\theta=1/\eta\sqrt{4N_1(N_2+1)+4N_2(N_1+1)}$. As $|N_1-N_2|\ll(N_1+N_2)/2$, the error scales at the HL with the total number of photons $N_1+N_2$.

\begin{figure}[t]
\leavevmode\includegraphics[width=1\columnwidth]{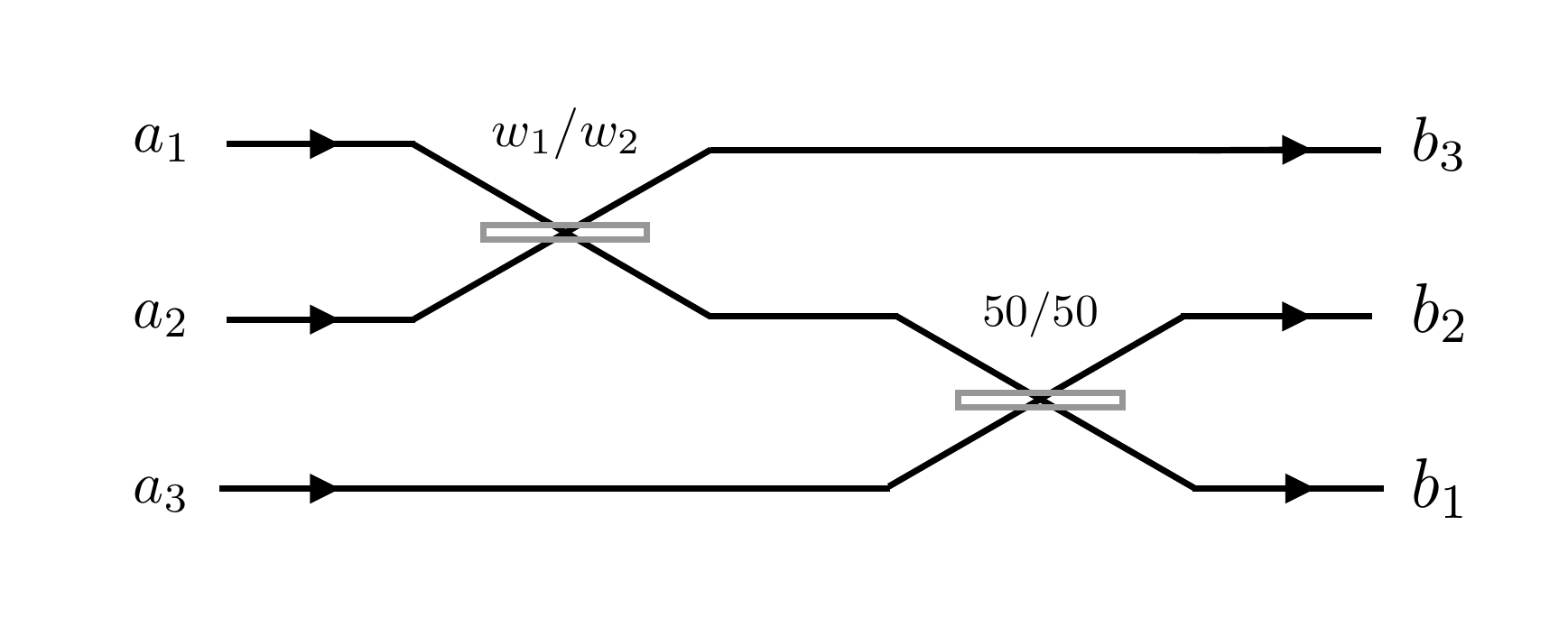}
\caption{Illustration of a simple linear network (consisting of only beam splitters) that implements the unitary $U$ described for the twin-Fock state approach to measuring $q$ with Heisenberg scaling.  Here we can measure $q=w_1\theta_1+w_2\theta_2$ for $w_1>0$ and $w_2>0$ with precision $1/\sqrt{2n(n+1)}$, using only a single reference port. The input state is $\ket{n,n,0}$ and the ratio above each beam-splitter is the transmission/reflection rate. The "classical" estimation strategy for measuring $q$ yields the precision $1/\sqrt{n(n+2})$, a factor of $\sqrt{2}$ larger for large $n$, using the same total number of photons.} 
\label{fig2} 
\end{figure} 

\emph{A general bound for arbitrary separable states}.---Equation \eqref{eq:bound} was derived assuming Fock-state inputs; here we show that a similar bound can be derived for arbitrary separable input states [\eref{eq:ngt} below].  Though this more general bound depends on different specific properties of the input states, it shares with \eref{eq:bound} the important characteristic that for fixed \emph{local} resources, i.e.\ if the properties of each input state [specifically $\langle n^2\rangle_{\rm max}$ in \eref{eq:ngt}] does not scale with $d$, then the minimal value of $\Delta q$ obeys the classical scaling $\sim 1/\sqrt{d}$.  

For an arbitrary separable input state $\rho$, the correct way to calculate the Fisher information depends on whether external phase references are assumed to be available in the measurement protocol~\cite{PhysRevA.85.011801}.  Here we make no such assumption, and therefore the Fisher information should be computed with respect to a phase averaged state,
\begin{align}
\varrho=\frac{1}{2\pi}\int_{0}^{2\pi}d\theta \exp(i\theta \hat{n})\rho\exp(-i\theta \hat{n}),
\end{align}
where $\hat{n}=\sum_{j}\hat{n}_j$ and $\theta$ is a global phase. However, because the Fisher information is convex and we need only obtain an upper bound, it is sufficient to calculate an upper bound on $\mathcal{F}$ for an arbitrary pure product state $\ket{\Psi}$, from which we can infer an upper bound on $\mathcal{F}$ for the separable density matrix $\varrho$.  Deriving a bound for an arbitrary pure state is still rather complex, and we defer a detailed analysis to the supplemental material.  Here, we simply quote the final result,
\begin{align} 
\label{eq:gBound}
\mathcal{F}_{\bm{w}} \leq \frac{A}{d}+B |\bm{w}|^2.
\end{align} 
We note that this bound is not tight (that is, it cannot necessarily be achieved). In Eq. \eqref{eq:gBound}, $A$ and $B$ depend only on moments of the input states (specifically $\braket{\hat{a}_j}$, $\braket{\hat{a}^{\dagger}_j\hat{a}^{\phantom\dagger}_j}$, $\braket{\hat{a}_j\hat{a}_j}$, $\braket{\hat{a}^{\dagger}_j\hat{a}^{\dagger}_j\hat{a}^{\phantom\dagger}_j}$, and $\braket{\hat{a}^{\dagger}_j\hat{a}^{\dagger}_j\hat{a}^{\phantom\dagger}_j\hat{a}^{\phantom\dagger}_j}$). Their exact form can be found in the supplement, but for our purposes it suffices only to know that they obey the bound $A + B< C^2\max_j m_j$, with $m_j\equiv\braket{(a_j^{\dagger}a_j)^2}$ and $C=20$.    Recalling that for well distributed weights we have $|\bm{w}|^2 \sim 1/d$, plugging \eref{eq:gBound} into \eref{eq:q_bound}, and defining $\langle n^2 \rangle_{\ms{max}} \equiv \max_j m_j$, we obtain   
\begin{equation} 
\label{eq:ngt}
   \Delta q \ge \frac{1}{C   \sqrt{d  \langle n^2 \rangle_{\ms{max}}}} . 
\end{equation}
If the resources are constrained locally rather than globally, such that $\langle n^2 \rangle_{\ms{max}}$ is independent of $d$, it follows that quantum metrology using linear-optical networks and separable input states cannot improve upon the classical scaling $\sim 1/\sqrt{d}$.

In addition to elucidating the resource requirements for quantum metrology with linear networks, the above results are also interesting from the point of view of quantifying non-classicality~\cite{hillery1987nonclassical}. Linear networks have the ability to reversibly transform non-classical but unentangled states into entangled ones, thus providing a route to quantifying non-classicality using measures of entanglement~\cite{Asboth:05,Vogel:14, Killoran:2016aa}. Our results suggest that it may be useful to refine this approach by using more stringent, operational measures of entanglement such as the ability of the entangled state to realize quantum-enhanced metrology. Our results reveal that under this measure the non-classicality of single-mode Fock states would increase with increasing number of photons, and the non-classicality of multi-mode states would tend to increase as the total number of photons is packed into fewer modes. These considerations suggest that it may be interesting to investigate how a quantitative measure of non-classicality might be obtained via quantum metrology, and more generally to explore the classes of states that are preserved under linear networks.

%\editms{In this letter, we have presented the framework of quantum metrology in a linear network using nonclassical product states. While a large variety of complex entangled states can be created in a linear network, not all nonclassical states are useful for distributed quantum metrology. For the task of estimating a linear combination of parameters distributed at distant nodes, the estimation sensitivity can only achieve the HL when a few modes prepare the nonclassical states with resources that scale with the total amount of resources. Our results show that the metrological usefulness of $d+1$ Fock state is much less than the twin-Fock state with the same amount of total photons at a multiport interferometer. This suggests the nonclassicality of Fock state \cite{Vogel:14} may be nonlinear with the photon number in terms of usefulness for quantum metrology. } 

\acknowledgements{We thank A. Deshpande, B. Gard, E. Goldschmidt, M. Lewenstein, J. V. Porto, S. Rolston, and V. Tamma for helpful discussions. Z.E. is supported in part by the ARCS Foundation. Z. E. and A.V.G. acknowledges support by ARL CDQI, ARO MURI, NSF QIS, ARO, NSF PFC at JQI, and AFOSR. This material is based upon work supported by, or in part by, the U. S. Army Research Laboratory and the U. S. Army Research Office under contract/grant number 025989-001, and by the ASD(R\&E) as part of the QSEP.}

\bibliography{Quantum_Metrology_in_Multiport_Linear_Interferometer}

%merlin.mbs apsrev4-1.bst 2010-07-25 4.21a (PWD, AO, DPC) hacked
%Control: key (0)
%Control: author (72) initials jnrlst
%Control: editor formatted (1) identically to author
%Control: production of article title (-1) disabled
%Control: page (0) single
%Control: year (1) truncated
%Control: production of eprint (0) enabled
\begin{thebibliography}{37}%
\makeatletter
\providecommand \@ifxundefined [1]{%
 \@ifx{#1\undefined}
}%
\providecommand \@ifnum [1]{%
 \ifnum #1\expandafter \@firstoftwo
 \else \expandafter \@secondoftwo
 \fi
}%
\providecommand \@ifx [1]{%
 \ifx #1\expandafter \@firstoftwo
 \else \expandafter \@secondoftwo
 \fi
}%
\providecommand \natexlab [1]{#1}%
\providecommand \enquote  [1]{``#1''}%
\providecommand \bibnamefont  [1]{#1}%
\providecommand \bibfnamefont [1]{#1}%
\providecommand \citenamefont [1]{#1}%
\providecommand \href@noop [0]{\@secondoftwo}%
\providecommand \href [0]{\begingroup \@sanitize@url \@href}%
\providecommand \@href[1]{\@@startlink{#1}\@@href}%
\providecommand \@@href[1]{\endgroup#1\@@endlink}%
\providecommand \@sanitize@url [0]{\catcode `\\12\catcode `\$12\catcode
  `\&12\catcode `\#12\catcode `\^12\catcode `\_12\catcode `\%12\relax}%
\providecommand \@@startlink[1]{}%
\providecommand \@@endlink[0]{}%
\providecommand \url  [0]{\begingroup\@sanitize@url \@url }%
\providecommand \@url [1]{\endgroup\@href {#1}{\urlprefix }}%
\providecommand \urlprefix  [0]{URL }%
\providecommand \Eprint [0]{\href }%
\providecommand \doibase [0]{http://dx.doi.org/}%
\providecommand \selectlanguage [0]{\@gobble}%
\providecommand \bibinfo  [0]{\@secondoftwo}%
\providecommand \bibfield  [0]{\@secondoftwo}%
\providecommand \translation [1]{[#1]}%
\providecommand \BibitemOpen [0]{}%
\providecommand \bibitemStop [0]{}%
\providecommand \bibitemNoStop [0]{.\EOS\space}%
\providecommand \EOS [0]{\spacefactor3000\relax}%
\providecommand \BibitemShut  [1]{\csname bibitem#1\endcsname}%
\let\auto@bib@innerbib\@empty
%</preamble>
\bibitem [{\citenamefont {Caves}(1981)}]{caves1981quantum}%
  \BibitemOpen
  \bibfield  {author} {\bibinfo {author} {\bibfnamefont {C.~M.}\ \bibnamefont
  {Caves}},\ }\href {\doibase 10.1103/PhysRevD.23.1693} {\bibfield  {journal}
  {\bibinfo  {journal} {Phys. Rev. D}\ }\textbf {\bibinfo {volume} {23}},\
  \bibinfo {pages} {1693} (\bibinfo {year} {1981})}\BibitemShut {NoStop}%
\bibitem [{\citenamefont {Holland}\ and\ \citenamefont
  {Burnett}(1993)}]{Holland:1993aa}%
  \BibitemOpen
  \bibfield  {author} {\bibinfo {author} {\bibfnamefont {M.}~\bibnamefont
  {Holland}}\ and\ \bibinfo {author} {\bibfnamefont {K.}~\bibnamefont
  {Burnett}},\ }\href {\doibase 10.1103/PhysRevLett.71.1355} {\bibfield
  {journal} {\bibinfo  {journal} {Phys. Rev. Lett.}\ }\textbf {\bibinfo
  {volume} {71}},\ \bibinfo {pages} {1355} (\bibinfo {year}
  {1993})}\BibitemShut {NoStop}%
\bibitem [{\citenamefont {Giovannetti}\ \emph {et~al.}(2004)\citenamefont
  {Giovannetti}, \citenamefont {Lloyd},\ and\ \citenamefont
  {Maccone}}]{giovannetti2004quantum}%
  \BibitemOpen
  \bibfield  {author} {\bibinfo {author} {\bibfnamefont {V.}~\bibnamefont
  {Giovannetti}}, \bibinfo {author} {\bibfnamefont {S.}~\bibnamefont {Lloyd}},
  \ and\ \bibinfo {author} {\bibfnamefont {L.}~\bibnamefont {Maccone}},\ }\href
  {http://science.sciencemag.org/content/306/5700/1330} {\bibfield  {journal}
  {\bibinfo  {journal} {Science}\ }\textbf {\bibinfo {volume} {306}},\ \bibinfo
  {pages} {1330} (\bibinfo {year} {2004})}\BibitemShut {NoStop}%
\bibitem [{\citenamefont {Giovannetti}\ \emph {et~al.}(2006)\citenamefont
  {Giovannetti}, \citenamefont {Lloyd},\ and\ \citenamefont
  {Maccone}}]{giovannetti2006quantum}%
  \BibitemOpen
  \bibfield  {author} {\bibinfo {author} {\bibfnamefont {V.}~\bibnamefont
  {Giovannetti}}, \bibinfo {author} {\bibfnamefont {S.}~\bibnamefont {Lloyd}},
  \ and\ \bibinfo {author} {\bibfnamefont {L.}~\bibnamefont {Maccone}},\ }\href
  {\doibase 10.1103/PhysRevLett.96.010401} {\bibfield  {journal} {\bibinfo
  {journal} {Phys. Rev. Lett.}\ }\textbf {\bibinfo {volume} {96}},\ \bibinfo
  {pages} {010401} (\bibinfo {year} {2006})}\BibitemShut {NoStop}%
\bibitem [{\citenamefont {Campos}\ \emph {et~al.}(2003)\citenamefont {Campos},
  \citenamefont {Gerry},\ and\ \citenamefont {Benmoussa}}]{campos2003optical}%
  \BibitemOpen
  \bibfield  {author} {\bibinfo {author} {\bibfnamefont {R.}~\bibnamefont
  {Campos}}, \bibinfo {author} {\bibfnamefont {C.~C.}\ \bibnamefont {Gerry}}, \
  and\ \bibinfo {author} {\bibfnamefont {A.}~\bibnamefont {Benmoussa}},\ }\href
  {\doibase 10.1103/PhysRevA.68.023810} {\bibfield  {journal} {\bibinfo
  {journal} {Phys. Rev. A}\ }\textbf {\bibinfo {volume} {68}},\ \bibinfo
  {pages} {023810} (\bibinfo {year} {2003})}\BibitemShut {NoStop}%
\bibitem [{\citenamefont {Pezz{\'e}}\ and\ \citenamefont
  {Smerzi}(2008)}]{pezze2008mach}%
  \BibitemOpen
  \bibfield  {author} {\bibinfo {author} {\bibfnamefont {L.}~\bibnamefont
  {Pezz{\'e}}}\ and\ \bibinfo {author} {\bibfnamefont {A.}~\bibnamefont
  {Smerzi}},\ }\href {\doibase 10.1103/PhysRevLett.100.073601} {\bibfield
  {journal} {\bibinfo  {journal} {Phys. Rev. Lett.}\ }\textbf {\bibinfo
  {volume} {100}},\ \bibinfo {pages} {073601} (\bibinfo {year}
  {2008})}\BibitemShut {NoStop}%
\bibitem [{\citenamefont {Cooper}\ \emph {et~al.}(2010)\citenamefont {Cooper},
  \citenamefont {Hallwood},\ and\ \citenamefont
  {Dunningham}}]{cooper2010entanglement}%
  \BibitemOpen
  \bibfield  {author} {\bibinfo {author} {\bibfnamefont {J.}~\bibnamefont
  {Cooper}}, \bibinfo {author} {\bibfnamefont {D.}~\bibnamefont {Hallwood}}, \
  and\ \bibinfo {author} {\bibfnamefont {J.}~\bibnamefont {Dunningham}},\
  }\href {\doibase 10.1103/PhysRevA.81.043624} {\bibfield  {journal} {\bibinfo
  {journal} {Phys. Rev. A}\ }\textbf {\bibinfo {volume} {81}},\ \bibinfo
  {pages} {043624} (\bibinfo {year} {2010})}\BibitemShut {NoStop}%
\bibitem [{\citenamefont {Birrittella}\ \emph {et~al.}(2012)\citenamefont
  {Birrittella}, \citenamefont {Mimih},\ and\ \citenamefont
  {Gerry}}]{birrittella2012multiphoton}%
  \BibitemOpen
  \bibfield  {author} {\bibinfo {author} {\bibfnamefont {R.}~\bibnamefont
  {Birrittella}}, \bibinfo {author} {\bibfnamefont {J.}~\bibnamefont {Mimih}},
  \ and\ \bibinfo {author} {\bibfnamefont {C.~C.}\ \bibnamefont {Gerry}},\
  }\href {\doibase 10.1103/PhysRevA.86.063828} {\bibfield  {journal} {\bibinfo
  {journal} {Phys. Rev. A}\ }\textbf {\bibinfo {volume} {86}},\ \bibinfo
  {pages} {063828} (\bibinfo {year} {2012})}\BibitemShut {NoStop}%
\bibitem [{\citenamefont {Gagatsos}\ \emph {et~al.}(2016)\citenamefont
  {Gagatsos}, \citenamefont {Branford},\ and\ \citenamefont
  {Datta}}]{gagatsos2016gaussian}%
  \BibitemOpen
  \bibfield  {author} {\bibinfo {author} {\bibfnamefont {C.~N.}\ \bibnamefont
  {Gagatsos}}, \bibinfo {author} {\bibfnamefont {D.}~\bibnamefont {Branford}},
  \ and\ \bibinfo {author} {\bibfnamefont {A.}~\bibnamefont {Datta}},\ }\href
  {\doibase 10.1103/PhysRevA.94.042342} {\bibfield  {journal} {\bibinfo
  {journal} {Phys. Rev. A}\ }\textbf {\bibinfo {volume} {94}},\ \bibinfo
  {pages} {042342} (\bibinfo {year} {2016})}\BibitemShut {NoStop}%
\bibitem [{\citenamefont {Eldredge}\ \emph {et~al.}(2016)\citenamefont
  {Eldredge}, \citenamefont {Foss-Feig}, \citenamefont {Rolston},\ and\
  \citenamefont {Gorshkov}}]{Zach:16}%
  \BibitemOpen
  \bibfield  {author} {\bibinfo {author} {\bibfnamefont {Z.}~\bibnamefont
  {Eldredge}}, \bibinfo {author} {\bibfnamefont {M.}~\bibnamefont {Foss-Feig}},
  \bibinfo {author} {\bibfnamefont {S.~L.}\ \bibnamefont {Rolston}}, \ and\
  \bibinfo {author} {\bibfnamefont {A.~V.}\ \bibnamefont {Gorshkov}},\ }\href
  {https://arxiv.org/abs/1607.04646} {\bibfield  {journal} {\bibinfo  {journal}
  {arXiv preprint arXiv:1607.04646}\ } (\bibinfo {year} {2016})}\BibitemShut
  {NoStop}%
\bibitem [{\citenamefont {Proctor}\ \emph {et~al.}(2017)\citenamefont
  {Proctor}, \citenamefont {Knott},\ and\ \citenamefont
  {Dunningham}}]{proctor2017networked}%
  \BibitemOpen
  \bibfield  {author} {\bibinfo {author} {\bibfnamefont {T.}~\bibnamefont
  {Proctor}}, \bibinfo {author} {\bibfnamefont {P.}~\bibnamefont {Knott}}, \
  and\ \bibinfo {author} {\bibfnamefont {J.}~\bibnamefont {Dunningham}},\
  }\href {https://arxiv.org/abs/1702.04271} {\bibfield  {journal} {\bibinfo
  {journal} {arXiv preprint arXiv:1702.04271}\ } (\bibinfo {year}
  {2017})}\BibitemShut {NoStop}%
\bibitem [{\citenamefont {Kim}\ \emph {et~al.}(2002)\citenamefont {Kim},
  \citenamefont {Son}, \citenamefont {Bu\ifmmode~\check{z}\else \v{z}\fi{}ek},\
  and\ \citenamefont {Knight}}]{Kim:02}%
  \BibitemOpen
  \bibfield  {author} {\bibinfo {author} {\bibfnamefont {M.~S.}\ \bibnamefont
  {Kim}}, \bibinfo {author} {\bibfnamefont {W.}~\bibnamefont {Son}}, \bibinfo
  {author} {\bibfnamefont {V.}~\bibnamefont {Bu\ifmmode~\check{z}\else
  \v{z}\fi{}ek}}, \ and\ \bibinfo {author} {\bibfnamefont {P.~L.}\ \bibnamefont
  {Knight}},\ }\href {\doibase 10.1103/PhysRevA.65.032323} {\bibfield
  {journal} {\bibinfo  {journal} {Phys. Rev. A}\ }\textbf {\bibinfo {volume}
  {65}},\ \bibinfo {pages} {032323} (\bibinfo {year} {2002})}\BibitemShut
  {NoStop}%
\bibitem [{\citenamefont {Xiang-bin}(2002)}]{Wang:02}%
  \BibitemOpen
  \bibfield  {author} {\bibinfo {author} {\bibfnamefont {W.}~\bibnamefont
  {Xiang-bin}},\ }\href {\doibase 10.1103/PhysRevA.66.024303} {\bibfield
  {journal} {\bibinfo  {journal} {Phys. Rev. A}\ }\textbf {\bibinfo {volume}
  {66}},\ \bibinfo {pages} {024303} (\bibinfo {year} {2002})}\BibitemShut
  {NoStop}%
\bibitem [{\citenamefont {Wolf}\ \emph {et~al.}(2003)\citenamefont {Wolf},
  \citenamefont {Eisert},\ and\ \citenamefont {Plenio}}]{Wolf:03}%
  \BibitemOpen
  \bibfield  {author} {\bibinfo {author} {\bibfnamefont {M.~M.}\ \bibnamefont
  {Wolf}}, \bibinfo {author} {\bibfnamefont {J.}~\bibnamefont {Eisert}}, \ and\
  \bibinfo {author} {\bibfnamefont {M.~B.}\ \bibnamefont {Plenio}},\ }\href
  {\doibase 10.1103/PhysRevLett.90.047904} {\bibfield  {journal} {\bibinfo
  {journal} {Phys. Rev. Lett.}\ }\textbf {\bibinfo {volume} {90}},\ \bibinfo
  {pages} {047904} (\bibinfo {year} {2003})}\BibitemShut {NoStop}%
\bibitem [{\citenamefont {Asb\'oth}\ \emph {et~al.}(2005)\citenamefont
  {Asb\'oth}, \citenamefont {Calsamiglia},\ and\ \citenamefont
  {Ritsch}}]{Asboth:05}%
  \BibitemOpen
  \bibfield  {author} {\bibinfo {author} {\bibfnamefont {J.~K.}\ \bibnamefont
  {Asb\'oth}}, \bibinfo {author} {\bibfnamefont {J.}~\bibnamefont
  {Calsamiglia}}, \ and\ \bibinfo {author} {\bibfnamefont {H.}~\bibnamefont
  {Ritsch}},\ }\href {\doibase 10.1103/PhysRevLett.94.173602} {\bibfield
  {journal} {\bibinfo  {journal} {Phys. Rev. Lett.}\ }\textbf {\bibinfo
  {volume} {94}},\ \bibinfo {pages} {173602} (\bibinfo {year}
  {2005})}\BibitemShut {NoStop}%
\bibitem [{\citenamefont {Vogel}\ and\ \citenamefont
  {Sperling}(2014)}]{Vogel:14}%
  \BibitemOpen
  \bibfield  {author} {\bibinfo {author} {\bibfnamefont {W.}~\bibnamefont
  {Vogel}}\ and\ \bibinfo {author} {\bibfnamefont {J.}~\bibnamefont
  {Sperling}},\ }\href {\doibase 10.1103/PhysRevA.89.052302} {\bibfield
  {journal} {\bibinfo  {journal} {Phys. Rev. A}\ }\textbf {\bibinfo {volume}
  {89}},\ \bibinfo {pages} {052302} (\bibinfo {year} {2014})}\BibitemShut
  {NoStop}%
\bibitem [{\citenamefont {Ge}\ \emph {et~al.}(2015)\citenamefont {Ge},
  \citenamefont {Tasgin},\ and\ \citenamefont {Zubairy}}]{ge2015conservation}%
  \BibitemOpen
  \bibfield  {author} {\bibinfo {author} {\bibfnamefont {W.}~\bibnamefont
  {Ge}}, \bibinfo {author} {\bibfnamefont {M.~E.}\ \bibnamefont {Tasgin}}, \
  and\ \bibinfo {author} {\bibfnamefont {M.~S.}\ \bibnamefont {Zubairy}},\
  }\href {\doibase 10.1103/PhysRevA.92.052328} {\bibfield  {journal} {\bibinfo
  {journal} {Phys. Rev. A}\ }\textbf {\bibinfo {volume} {92}},\ \bibinfo
  {pages} {052328} (\bibinfo {year} {2015})}\BibitemShut {NoStop}%
\bibitem [{\citenamefont {Killoran}\ \emph {et~al.}(2016)\citenamefont
  {Killoran}, \citenamefont {Steinhoff},\ and\ \citenamefont
  {Plenio}}]{Killoran:2016aa}%
  \BibitemOpen
  \bibfield  {author} {\bibinfo {author} {\bibfnamefont {N.}~\bibnamefont
  {Killoran}}, \bibinfo {author} {\bibfnamefont {F.~E.}\ \bibnamefont
  {Steinhoff}}, \ and\ \bibinfo {author} {\bibfnamefont {M.~B.}\ \bibnamefont
  {Plenio}},\ }\href {\doibase 10.1103/PhysRevLett.116.080402} {\bibfield
  {journal} {\bibinfo  {journal} {Phys. Rev. Lett.}\ }\textbf {\bibinfo
  {volume} {116}},\ \bibinfo {pages} {080402} (\bibinfo {year}
  {2016})}\BibitemShut {NoStop}%
\bibitem [{\citenamefont {Aaronson}\ and\ \citenamefont
  {Arkhipov}(2013)}]{Aaronson_2013}%
  \BibitemOpen
  \bibfield  {author} {\bibinfo {author} {\bibfnamefont {S.}~\bibnamefont
  {Aaronson}}\ and\ \bibinfo {author} {\bibfnamefont {A.}~\bibnamefont
  {Arkhipov}},\ }\href {http://theoryofcomputing.org/articles/v009a004/}
  {\bibfield  {journal} {\bibinfo  {journal} {Theory of Computing}\ }\textbf
  {\bibinfo {volume} {9}},\ \bibinfo {pages} {143} (\bibinfo {year}
  {2013})}\BibitemShut {NoStop}%
\bibitem [{\citenamefont {Tillmann}\ \emph {et~al.}(2013)\citenamefont
  {Tillmann}, \citenamefont {Dakic}, \citenamefont {Heilmann}, \citenamefont
  {Nolte}, \citenamefont {Szameit},\ and\ \citenamefont
  {Walther}}]{Tillman_2013}%
  \BibitemOpen
  \bibfield  {author} {\bibinfo {author} {\bibfnamefont {M.}~\bibnamefont
  {Tillmann}}, \bibinfo {author} {\bibfnamefont {B.}~\bibnamefont {Dakic}},
  \bibinfo {author} {\bibfnamefont {R.}~\bibnamefont {Heilmann}}, \bibinfo
  {author} {\bibfnamefont {S.}~\bibnamefont {Nolte}}, \bibinfo {author}
  {\bibfnamefont {A.}~\bibnamefont {Szameit}}, \ and\ \bibinfo {author}
  {\bibfnamefont {P.}~\bibnamefont {Walther}},\ }\href
  {http://dx.doi.org/10.1038/nphoton.2013.102} {\bibfield  {journal} {\bibinfo
  {journal} {Nat. Photon}\ }\textbf {\bibinfo {volume} {7}},\ \bibinfo {pages}
  {540} (\bibinfo {year} {2013})}\BibitemShut {NoStop}%
\bibitem [{\citenamefont {Broome}\ \emph {et~al.}(2013)\citenamefont {Broome},
  \citenamefont {Fedrizzi}, \citenamefont {Rahimi-Keshari}, \citenamefont
  {Dove}, \citenamefont {Aaronson}, \citenamefont {Ralph},\ and\ \citenamefont
  {White}}]{Broome794}%
  \BibitemOpen
  \bibfield  {author} {\bibinfo {author} {\bibfnamefont {M.~A.}\ \bibnamefont
  {Broome}}, \bibinfo {author} {\bibfnamefont {A.}~\bibnamefont {Fedrizzi}},
  \bibinfo {author} {\bibfnamefont {S.}~\bibnamefont {Rahimi-Keshari}},
  \bibinfo {author} {\bibfnamefont {J.}~\bibnamefont {Dove}}, \bibinfo {author}
  {\bibfnamefont {S.}~\bibnamefont {Aaronson}}, \bibinfo {author}
  {\bibfnamefont {T.~C.}\ \bibnamefont {Ralph}}, \ and\ \bibinfo {author}
  {\bibfnamefont {A.~G.}\ \bibnamefont {White}},\ }\href {\doibase
  10.1126/science.1231440} {\bibfield  {journal} {\bibinfo  {journal}
  {Science}\ }\textbf {\bibinfo {volume} {339}},\ \bibinfo {pages} {794}
  (\bibinfo {year} {2013})}\BibitemShut {NoStop}%
\bibitem [{\citenamefont {Spring}\ \emph {et~al.}(2013)\citenamefont {Spring},
  \citenamefont {Metcalf}, \citenamefont {Humphreys}, \citenamefont
  {Kolthammer}, \citenamefont {Jin}, \citenamefont {Barbieri}, \citenamefont
  {Datta}, \citenamefont {Thomas-Peter}, \citenamefont {Langford},
  \citenamefont {Kundys}, \citenamefont {Gates}, \citenamefont {Smith},
  \citenamefont {Smith},\ and\ \citenamefont {Walmsley}}]{Spring798}%
  \BibitemOpen
  \bibfield  {author} {\bibinfo {author} {\bibfnamefont {J.~B.}\ \bibnamefont
  {Spring}}, \bibinfo {author} {\bibfnamefont {B.~J.}\ \bibnamefont {Metcalf}},
  \bibinfo {author} {\bibfnamefont {P.~C.}\ \bibnamefont {Humphreys}}, \bibinfo
  {author} {\bibfnamefont {W.~S.}\ \bibnamefont {Kolthammer}}, \bibinfo
  {author} {\bibfnamefont {X.-M.}\ \bibnamefont {Jin}}, \bibinfo {author}
  {\bibfnamefont {M.}~\bibnamefont {Barbieri}}, \bibinfo {author}
  {\bibfnamefont {A.}~\bibnamefont {Datta}}, \bibinfo {author} {\bibfnamefont
  {N.}~\bibnamefont {Thomas-Peter}}, \bibinfo {author} {\bibfnamefont {N.~K.}\
  \bibnamefont {Langford}}, \bibinfo {author} {\bibfnamefont {D.}~\bibnamefont
  {Kundys}}, \bibinfo {author} {\bibfnamefont {J.~C.}\ \bibnamefont {Gates}},
  \bibinfo {author} {\bibfnamefont {B.~J.}\ \bibnamefont {Smith}}, \bibinfo
  {author} {\bibfnamefont {P.~G.~R.}\ \bibnamefont {Smith}}, \ and\ \bibinfo
  {author} {\bibfnamefont {I.~A.}\ \bibnamefont {Walmsley}},\ }\href {\doibase
  10.1126/science.1231692} {\bibfield  {journal} {\bibinfo  {journal}
  {Science}\ }\textbf {\bibinfo {volume} {339}},\ \bibinfo {pages} {798}
  (\bibinfo {year} {2013})}\BibitemShut {NoStop}%
\bibitem [{\citenamefont {Crespi}\ \emph {et~al.}(2013)\citenamefont {Crespi},
  \citenamefont {Osellame}, \citenamefont {Ramponi}, \citenamefont {Brod},
  \citenamefont {Galvao}, \citenamefont {Spagnolo}, \citenamefont {Vitelli},
  \citenamefont {Maiorino}, \citenamefont {Mataloni},\ and\ \citenamefont
  {Sciarrino}}]{Crespi2013}%
  \BibitemOpen
  \bibfield  {author} {\bibinfo {author} {\bibfnamefont {A.}~\bibnamefont
  {Crespi}}, \bibinfo {author} {\bibfnamefont {R.}~\bibnamefont {Osellame}},
  \bibinfo {author} {\bibfnamefont {R.}~\bibnamefont {Ramponi}}, \bibinfo
  {author} {\bibfnamefont {D.~J.}\ \bibnamefont {Brod}}, \bibinfo {author}
  {\bibfnamefont {E.~F.}\ \bibnamefont {Galvao}}, \bibinfo {author}
  {\bibfnamefont {N.}~\bibnamefont {Spagnolo}}, \bibinfo {author}
  {\bibfnamefont {C.}~\bibnamefont {Vitelli}}, \bibinfo {author} {\bibfnamefont
  {E.}~\bibnamefont {Maiorino}}, \bibinfo {author} {\bibfnamefont
  {P.}~\bibnamefont {Mataloni}}, \ and\ \bibinfo {author} {\bibfnamefont
  {F.}~\bibnamefont {Sciarrino}},\ }\href
  {http://dx.doi.org/10.1038/nphoton.2013.112} {\bibfield  {journal} {\bibinfo
  {journal} {Nat. Photon}\ }\textbf {\bibinfo {volume} {7}},\ \bibinfo {pages}
  {545} (\bibinfo {year} {2013})}\BibitemShut {NoStop}%
\bibitem [{\citenamefont {Kok}\ and\ \citenamefont
  {Braunstein}(2000)}]{kok2000limitations}%
  \BibitemOpen
  \bibfield  {author} {\bibinfo {author} {\bibfnamefont {P.}~\bibnamefont
  {Kok}}\ and\ \bibinfo {author} {\bibfnamefont {S.~L.}\ \bibnamefont
  {Braunstein}},\ }\href {\doibase 10.1103/PhysRevA.62.064301} {\bibfield
  {journal} {\bibinfo  {journal} {Phys. Rev. A}\ }\textbf {\bibinfo {volume}
  {62}},\ \bibinfo {pages} {064301} (\bibinfo {year} {2000})}\BibitemShut
  {NoStop}%
\bibitem [{\citenamefont {Kok}\ \emph {et~al.}(2007)\citenamefont {Kok},
  \citenamefont {Munro}, \citenamefont {Nemoto}, \citenamefont {Ralph},
  \citenamefont {Dowling},\ and\ \citenamefont {Milburn}}]{Kok:2007aa}%
  \BibitemOpen
  \bibfield  {author} {\bibinfo {author} {\bibfnamefont {P.}~\bibnamefont
  {Kok}}, \bibinfo {author} {\bibfnamefont {W.~J.}\ \bibnamefont {Munro}},
  \bibinfo {author} {\bibfnamefont {K.}~\bibnamefont {Nemoto}}, \bibinfo
  {author} {\bibfnamefont {T.~C.}\ \bibnamefont {Ralph}}, \bibinfo {author}
  {\bibfnamefont {J.~P.}\ \bibnamefont {Dowling}}, \ and\ \bibinfo {author}
  {\bibfnamefont {G.~J.}\ \bibnamefont {Milburn}},\ }\href {\doibase
  10.1103/RevModPhys.79.135} {\bibfield  {journal} {\bibinfo  {journal}
  {Reviews of Modern Physics}\ }\textbf {\bibinfo {volume} {79}},\ \bibinfo
  {pages} {135} (\bibinfo {year} {2007})}\BibitemShut {NoStop}%
\bibitem [{Note1()}]{Note1}%
  \BibitemOpen
  \bibinfo {note} {Note that by classical we imply only a lack of correlations
  between the nodes, while the estimation procedure employed at each node is
  \protect \emph {not} assumed to be classical. The natural definition of
  ``classical'' might be augmented to allow classical distribution of the
  initial resources, but this will not improve the scaling of $\Delta q
  _{\protect \rm cl}$ for well-distributed $\protect \bm {w}$.}\BibitemShut
  {Stop}%
\bibitem [{Note2()}]{Note2}%
  \BibitemOpen
  \bibinfo {note} {This inequality is short-hand for the asymptotic per-trial
  covariance. Strictly speaking, for $\nu $ trials of an estimation procedure
  we have ${\protect \rm Cov}(\protect \bm {\Theta })\geq \left (\nu \protect
  \mathcal {F}\right )^{-1}$, which is saturable in the large $\nu $
  limit.}\BibitemShut {Stop}%
\bibitem [{\citenamefont {Paris}(2009)}]{Paris_2009}%
  \BibitemOpen
  \bibfield  {author} {\bibinfo {author} {\bibfnamefont {M.~G.~A.}\
  \bibnamefont {Paris}},\ }\href
  {http://www.worldscientific.com/doi/abs/10.1142/S0219749909004839} {\bibfield
   {journal} {\bibinfo  {journal} {Int. J. Quant. Inf.}\ }\textbf {\bibinfo
  {volume} {7}},\ \bibinfo {pages} {125} (\bibinfo {year} {2009})}\BibitemShut
  {NoStop}%
\bibitem [{\citenamefont {Petz}(1994)}]{Petz_1994}%
  \BibitemOpen
  \bibfield  {author} {\bibinfo {author} {\bibfnamefont {D.}~\bibnamefont
  {Petz}},\ }\href {https://eudml.org/doc/262566} {\bibfield  {journal}
  {\bibinfo  {journal} {Banach Centre Publications}\ }\textbf {\bibinfo
  {volume} {30}},\ \bibinfo {pages} {287} (\bibinfo {year} {1994})}\BibitemShut
  {NoStop}%
\bibitem [{\citenamefont {Mirsky}(1975)}]{Mirsky_1975}%
  \BibitemOpen
  \bibfield  {author} {\bibinfo {author} {\bibfnamefont {L.}~\bibnamefont
  {Mirsky}},\ }\href {https://link.springer.com/article/10.1007/BF01647331}
  {\bibfield  {journal} {\bibinfo  {journal} {Mon. f. Math.}\ }\textbf
  {\bibinfo {volume} {79}},\ \bibinfo {pages} {303} (\bibinfo {year}
  {1975})}\BibitemShut {NoStop}%
\bibitem [{\citenamefont {Olson}\ \emph {et~al.}(2016)\citenamefont {Olson},
  \citenamefont {Motes}, \citenamefont {Birchall}, \citenamefont {Studer},
  \citenamefont {LaBorde}, \citenamefont {Moulder}, \citenamefont {Rohde},\
  and\ \citenamefont {Dowling}}]{Olson:2016}%
  \BibitemOpen
  \bibfield  {author} {\bibinfo {author} {\bibfnamefont {J.~P.}\ \bibnamefont
  {Olson}}, \bibinfo {author} {\bibfnamefont {K.~R.}\ \bibnamefont {Motes}},
  \bibinfo {author} {\bibfnamefont {P.~M.}\ \bibnamefont {Birchall}}, \bibinfo
  {author} {\bibfnamefont {N.~M.}\ \bibnamefont {Studer}}, \bibinfo {author}
  {\bibfnamefont {M.}~\bibnamefont {LaBorde}}, \bibinfo {author} {\bibfnamefont
  {T.}~\bibnamefont {Moulder}}, \bibinfo {author} {\bibfnamefont {P.~P.}\
  \bibnamefont {Rohde}}, \ and\ \bibinfo {author} {\bibfnamefont {J.~P.}\
  \bibnamefont {Dowling}},\ }\href {https://arxiv.org/abs/1610.07128}
  {\bibfield  {journal} {\bibinfo  {journal} {arXiv preprint arXiv:1610.07128}\
  } (\bibinfo {year} {2016})}\BibitemShut {NoStop}%
\bibitem [{\citenamefont {Cooper}\ \emph {et~al.}(2013)\citenamefont {Cooper},
  \citenamefont {Wright}, \citenamefont {S{\"o}ller},\ and\ \citenamefont
  {Smith}}]{cooper2013experimental}%
  \BibitemOpen
  \bibfield  {author} {\bibinfo {author} {\bibfnamefont {M.}~\bibnamefont
  {Cooper}}, \bibinfo {author} {\bibfnamefont {L.~J.}\ \bibnamefont {Wright}},
  \bibinfo {author} {\bibfnamefont {C.}~\bibnamefont {S{\"o}ller}}, \ and\
  \bibinfo {author} {\bibfnamefont {B.~J.}\ \bibnamefont {Smith}},\ }\href
  {\doibase 10.1364/OE.21.005309} {\bibfield  {journal} {\bibinfo  {journal}
  {Optics Express}\ }\textbf {\bibinfo {volume} {21}},\ \bibinfo {pages} {5309}
  (\bibinfo {year} {2013})}\BibitemShut {NoStop}%
\bibitem [{\citenamefont {Reck}\ \emph {et~al.}(1994)\citenamefont {Reck},
  \citenamefont {Zeilinger}, \citenamefont {Bernstein},\ and\ \citenamefont
  {Bertani}}]{Reck:1994aa}%
  \BibitemOpen
  \bibfield  {author} {\bibinfo {author} {\bibfnamefont {M.}~\bibnamefont
  {Reck}}, \bibinfo {author} {\bibfnamefont {A.}~\bibnamefont {Zeilinger}},
  \bibinfo {author} {\bibfnamefont {H.~J.}\ \bibnamefont {Bernstein}}, \ and\
  \bibinfo {author} {\bibfnamefont {P.}~\bibnamefont {Bertani}},\ }\href
  {\doibase 10.1103/PhysRevLett.73.58} {\bibfield  {journal} {\bibinfo
  {journal} {Phys. Rev. Lett.}\ }\textbf {\bibinfo {volume} {73}},\ \bibinfo
  {pages} {58} (\bibinfo {year} {1994})}\BibitemShut {NoStop}%
\bibitem [{\citenamefont {\ifmmode~\dot{Z}\else \.{Z}\fi{}ukowski}\ \emph
  {et~al.}(1997)\citenamefont {\ifmmode~\dot{Z}\else \.{Z}\fi{}ukowski},
  \citenamefont {Zeilinger},\ and\ \citenamefont
  {Horne}}]{marek1997realizable}%
  \BibitemOpen
  \bibfield  {author} {\bibinfo {author} {\bibfnamefont {M.}~\bibnamefont
  {\ifmmode~\dot{Z}\else \.{Z}\fi{}ukowski}}, \bibinfo {author} {\bibfnamefont
  {A.}~\bibnamefont {Zeilinger}}, \ and\ \bibinfo {author} {\bibfnamefont
  {M.~A.}\ \bibnamefont {Horne}},\ }\href {\doibase 10.1103/PhysRevA.55.2564}
  {\bibfield  {journal} {\bibinfo  {journal} {Phys. Rev. A}\ }\textbf {\bibinfo
  {volume} {55}},\ \bibinfo {pages} {2564} (\bibinfo {year}
  {1997})}\BibitemShut {NoStop}%
\bibitem [{\citenamefont {Kardyna{\l}}\ \emph {et~al.}(2008)\citenamefont
  {Kardyna{\l}}, \citenamefont {Yuan},\ and\ \citenamefont
  {Shields}}]{kardynal2008avalanche}%
  \BibitemOpen
  \bibfield  {author} {\bibinfo {author} {\bibfnamefont {B.}~\bibnamefont
  {Kardyna{\l}}}, \bibinfo {author} {\bibfnamefont {Z.}~\bibnamefont {Yuan}}, \
  and\ \bibinfo {author} {\bibfnamefont {A.}~\bibnamefont {Shields}},\ }\href
  {http://www.nature.com/nphoton/journal/v2/n7/abs/nphoton.2008.101.html}
  {\bibfield  {journal} {\bibinfo  {journal} {Nature Photonics}\ }\textbf
  {\bibinfo {volume} {2}},\ \bibinfo {pages} {425} (\bibinfo {year}
  {2008})}\BibitemShut {NoStop}%
\bibitem [{\citenamefont {Jarzyna}\ and\ \citenamefont
  {Demkowicz-Dobrza\ifmmode~\acute{n}\else
  \'{n}\fi{}ski}(2012)}]{PhysRevA.85.011801}%
  \BibitemOpen
  \bibfield  {author} {\bibinfo {author} {\bibfnamefont {M.}~\bibnamefont
  {Jarzyna}}\ and\ \bibinfo {author} {\bibfnamefont {R.}~\bibnamefont
  {Demkowicz-Dobrza\ifmmode~\acute{n}\else \'{n}\fi{}ski}},\ }\href {\doibase
  10.1103/PhysRevA.85.011801} {\bibfield  {journal} {\bibinfo  {journal} {Phys.
  Rev. A}\ }\textbf {\bibinfo {volume} {85}},\ \bibinfo {pages} {011801}
  (\bibinfo {year} {2012})}\BibitemShut {NoStop}%
\bibitem [{\citenamefont {Hillery}(1987)}]{hillery1987nonclassical}%
  \BibitemOpen
  \bibfield  {author} {\bibinfo {author} {\bibfnamefont {M.}~\bibnamefont
  {Hillery}},\ }\href {\doibase 10.1103/PhysRevA.35.725} {\bibfield  {journal}
  {\bibinfo  {journal} {Phys. Rev. A}\ }\textbf {\bibinfo {volume} {35}},\
  \bibinfo {pages} {725} (\bibinfo {year} {1987})}\BibitemShut {NoStop}%
\end{thebibliography}%

\begin{widetext}
\title{}
\maketitle

\newpage
\appendix\newpage\markboth{Appendix}{}
\setcounter{equation}{0}
\renewcommand{\thesection}{S}
\numberwithin{equation}{section}

\section{Supplementary Materials}

\section{S1. Quantum Fisher information for an arbitrary separable state}

The quantum Fisher information matrix element for an arbitrary input state $\ket{\Psi}$ under the unitary transformation $U$ is given by
\begin{align}
\mathcal{F}_{jk}=4\bra{\Psi_U}\hat{n}_j\hat{n}_k\ket{\Psi_U}  - 4\bra{\Psi_U}\hat{n}_j\ket{\Psi_U}\bra{\Psi_U}\hat{n}_k\ket{\Psi_U},
\end{align}
where $\hat{n}_j=b_j^{\dagger}b_j$ are the photon number operators in the output mode. The unitary transformation $U$ relates the output mode to the input mode as $b_j^{\dagger}=\sum_kU_{jk}a^{\dagger}_k$. Therefore, the quantum Fisher information written in terms of the input modes and the initial input state is given by
\begin{eqnarray}
\mathcal{F}_{jk}&=&4\bra{\Psi}\sum_{l,m,r,s}U_{jl}a^{\dagger}_lU_{jm}^{\ast}a_mU_{kr}a^{\dagger}_rU_{ks}^{\ast}a_s\ket{\Psi}  -4 \bra{\Psi}\sum_{l,m}U_{jl}a^{\dagger}_lU_{jm}^{\ast}a_m\ket{\Psi}\bra{\Psi}\sum_{r,s}U_{kr}a^{\dagger}_rU_{ks}^{\ast}a_s\ket{\Psi}\nonumber\\
&=&4\sum_{l,m,r,s}U^{}_{jl}U^{*}_{jm}U^{}_{kr}U^{*}_{ks}\Big(\langle\hat{a}^{\dagger}_l\hat{a}^{}_m\hat{a}^{\dagger}_r\hat{a}^{}_s\rangle-\langle\hat{a}^{\dagger}_l\hat{a}^{}_m\rangle\langle\hat{a}^{\dagger}_r\hat{a}^{}_s\rangle\Big),
\end{eqnarray}
which is the expression in Eq. \eqref{eq:FQFI_exp}. The expression on $\mathcal{F}_{\bm{w}}$ is then given by
\begin{eqnarray}
\mathcal{F}_{\bm{w}}=4\sum_{j,k}w_jw_k\sum_{l,m,r,s}U^{}_{jl}U^{*}_{jm}U^{}_{kr}U^{*}_{ks}\Big(\langle\hat{a}^{\dagger}_l\hat{a}^{}_m\hat{a}^{\dagger}_r\hat{a}^{}_s\rangle-\langle\hat{a}^{\dagger}_l\hat{a}^{}_m\rangle\langle\hat{a}^{\dagger}_r\hat{a}^{}_s\rangle\Big),
\end{eqnarray}

To simplify this expression for a separable input state, we categorize the summation $\sum_{l,m,r,s}$ into different parts depending on how many indices are identical. Hence it is expanded as
\begin{eqnarray}
\mathcal{F}_{\bm{w}}&=&4\sum_{j,k}w_jw_k\left[\sum_{l}|U_{jl}|^2|U_{kl}|^2\left(\braket{a^{\dagger}_la_la^{\dagger}_la_l}-\braket{a^{\dagger}_la_l}^2\right)\right.  \nonumber\\
&+&\sum_{l\ne s} |U_{jl}|^2U_{kl}U_{ks}^{\ast}\left(\braket{a^{\dagger}_la_la^{\dagger}_l}\braket{a_s}+\braket{a^{\dagger}_la^{\dagger}_la_l}\braket{a_s}-2\braket{a^{\dagger}_la_l}\braket{a^{\dagger}_l}\braket{a_s}\right)+c.c.\nonumber\\
&+&\sum_{l\ne m}U_{jl}U^{\ast}_{jm}U_{kl}U^{\ast}_{km}\left(\braket{a^{\dagger2}_l}\braket{a_m^2}-\braket{a^{\dagger}_l}^2\braket{a_m}^2\right)\nonumber\\
&+&\sum_{l\ne m}U_{jl}U^{\ast}_{jm}U_{km}U^{\ast}_{kl}\left(\braket{a^{\dagger}_la_l}\braket{a_ma^{\dagger}_m}-\braket{a^{\dagger}_l}\braket{a_m}\braket{a^{\dagger}_m}\braket{a_l}\right)\nonumber\\
&+&\sum_{l\ne m\ne s}U_{jl}U^{\ast}_{jm}U_{kl}U^{\ast}_{ks}\left(\braket{a^{\dagger2}_l}\braket{a_m}\braket{a_s}-\braket{a^{\dagger}_l}^2\braket{a_m}\braket{a_s}\right)+c.c.\nonumber\\
&+&\left.\sum_{l\ne m\ne r}U_{jl}U^{\ast}_{jm}U_{kr}U^{\ast}_{kl}\left(\braket{a^{\dagger}_la_l}\braket{a_m}\braket{a^{\dagger}_r}+\braket{a_la^{\dagger}_l}\braket{a_m}\braket{a^{\dagger}_r}-2\braket{a^{\dagger}_l}\braket{a_m}\braket{a^{\dagger}_r}\braket{a_l}\right)\right],
\end{eqnarray}
where the terms for which $ {l,m}\bigcap{r,s}=\emptyset$ vanish.
Defining the single-mode moments for the input states as
\begin{align}
\label{eq:sm1}
\alpha_j &\equiv \braket{a_j}, & n_j &\equiv \braket{a^{\dagger}_ja_j}, & \xi_j & \equiv \braket{a^2_j}, & \beta_j & \equiv \braket{a^{\dagger}_ja_ja_j}, & m_j & \equiv \braket{(a^{\dagger}_ja_j)^2}, & v_j & \equiv m_j-n_j^2,
 \end{align}
we simplify $\mathcal{F}_{\bm{w}}$ as 
\begin{eqnarray}
\label{eq:Fw}
\mathcal{F}_{\bm{w}}/4&=&\sum_{l}\mathcal{S}_{ll}^2v_l+\sum_{l\ne m} \mathcal{S}_{ml}\mathcal{S}_{lm}\left[n_l(n_m+1)-|\alpha_l|^2|\alpha_m|^2\right]\nonumber\\
&+&\sum_{l\ne m} \mathcal{S}_{ml}^2\left(\xi_l^{\ast}\xi_m-\alpha_l^{\ast 2}\alpha_m^2\right)+\sum_{l\ne m\ne s}\mathcal{S}_{ml}\mathcal{S}_{ls}\left(2n_l+1-2|\alpha_l|^2\right)\alpha_m\alpha^{\ast}_s\nonumber\\
&+&\left[\sum_{l\ne m\ne s}\mathcal{S}_{ml}\mathcal{S}_{sl}\left(\xi_l^{\ast}-\alpha_l^{\ast 2}\right)\alpha_m\alpha_s+c.c\right]+\left[\sum_{l\ne s}\mathcal{S}_{ll}\mathcal{S}_{sl}\left(2\beta_l^{\ast}+\alpha_l^{\ast}-2n_l\alpha^{\ast}_l\right)\alpha_s+c.c.\right],\end{eqnarray}
where $\mathcal{S}_{lm}=\sum_{j}U^{}_{jm}w_jU^{*}_{jl}$. Note that here $n_j$, which was previously used to describe the photon number in a Fock state, is now an expectation value in a state without definite photon number. For Fock state inputs, we set $v_l=\beta_l=\xi_l=\alpha_l=0$ and we recover the result in Eq. \eqref{eq:F_reduce}. Defining the terms on the right-hand side of Eq. \eqref{eq:Fw} as $F_j$ $(j=1,2,\cdots, 6.)$ respectively, we have $\mathcal{F}_{\bm{w}}/4=F_1+F_2+\cdots+F_6$.

%with $|w_{\max}|=\max_l|w_l|$, $n_{\text{max}}=\max_l\left(n_l+1/2-|\alpha_l|^2\right)$, $\alpha_{\max}=\max_j|\alpha_j|$, $\xi_{\max}=\max_j\left|\xi_l^{\ast}-\alpha_l^{\ast 2}\right|$, $\beta_{\max}=\max_l\left|\beta_l^{\ast}+\alpha_l^{\ast}/2-n_l\alpha^{\ast}_l\right|$, $v_{\text{max}}=\max_lv_l$, $M_{\text{max}}=\max\limits_{l\ne m}\left[n_l(n_m+1)-|\alpha_l|^2|\alpha_m|^2\right]$, $\Xi_{\text{max}}=\max\limits_{l\ne m}\left|\xi_l^{\ast}\xi_m-\alpha_l^{\ast 2}\alpha_m^2\right|$ and the 
\section{S2. An upper bound on quantum Fisher information}
Now we are going to derive the upper bound in Eq. \eqref{eq:gBound} for an arbitrary separable input state. We proceed by adding positive terms to each term in the above equation. The first term is 
\begin{eqnarray}
F_1&&=\sum_{l}\mathcal{S}_{ll}^2v_l \le v_{\text{max}}\sum_{l}\mathcal{S}_{ll}^2\le v_{\text{max}}\left(\sum_{l}\mathcal{S}_{ll}^2+\sum_{l\ne m}\mathcal{S}_{lm}\mathcal{S}_{ml}\right)\nonumber\\
&&=v_{\text{max}}\sum_{l, m}\mathcal{S}_{lm}\mathcal{S}_{ml}=v_{\text{max}}\sum_l w_l^2,
\end{eqnarray}
where the first inequality holds since $v_{\text{max}}=\max\limits_lv_l$, the second inequality holds because $\mathcal{S}_{lm}\mathcal{S}_{ml}=\left|\mathcal{S}_{lm}\right|^2\ge0$ and the last equal sign is due to $\sum_{l, m}\mathcal{S}_{lm}\mathcal{S}_{ml}=\sum_l w_l^2$. Similarly, the second term is 
\begin{eqnarray}
F_2&=&\sum_{l\ne m} \mathcal{S}_{ml}\mathcal{S}_{lm}\left[n_l(n_m+1)-|\alpha_l|^2|\alpha_m|^2\right]\le M_{\text{max}}\sum_{l\ne m} \mathcal{S}_{ml}\mathcal{S}_{lm}\nonumber\\
&\le& M_{\text{max}}\sum_{l,m} \mathcal{S}_{ml}\mathcal{S}_{lm}=M_{\text{max}}\sum_l w_l^2,
\end{eqnarray}
where $M_{\text{max}}=\max\limits_{l\ne m}\left[n_l(n_m+1)-|\alpha_l|^2|\alpha_m|^2\right]$. We obtain an inequality on the third term as
\begin{eqnarray}
F_3&=&\sum_{l\ne m} \mathcal{S}_{ml}^2\left(\xi_l^{\ast}\xi_m-\alpha_l^{\ast 2}\alpha_m^2\right)\le \sum_{l\ne m} \left|\mathcal{S}_{ml}^2\left(\xi_l^{\ast}\xi_m-\alpha_l^{\ast 2}\alpha_m^2\right)\right|\nonumber\\
&=&\sum_{l\ne m}\mathcal{S}_{ml}\mathcal{S}_{lm}\left|\xi_l^{\ast}\xi_m-\alpha_l^{\ast 2}\alpha_m^2\right|\le\Xi_{\text{max}}\sum_{l,m}\mathcal{S}_{ml}\mathcal{S}_{lm}=\Xi_{\text{max}}\sum_l w_l^2,
\label{eq:S9}
\end{eqnarray}
where $\Xi_{\text{max}}=\max\limits_{l\ne m}\left|\xi_l^{\ast}\xi_m-\alpha_l^{\ast 2}\alpha_m^2\right|$. The steps to get an upper bound on the fourth term to the sixth term are more involved. First, we expand the fourth term in $\mathcal{F}_{\bm{w}}/4$ as
\begin{eqnarray}
F_4&=&\sum_{l,m,s}\mathcal{S}_{ml}\mathcal{S}_{ls}\left(2n_l+1-2|\alpha_l|^2\right)\alpha_m\alpha^{\ast}_s-\sum_{l}\mathcal{S}_{ll}^2\left(2n_l+1-2|\alpha_l|^2\right)|\alpha_l|^2\nonumber\\
&-&\sum_{l\ne m}\mathcal{S}_{ml}\mathcal{S}_{lm}\left(2n_l+1-2|\alpha_l|^2\right)|\alpha_m|^2-\left[\sum_{l\ne m}\mathcal{S}_{ml}\mathcal{S}_{ll}\left(2n_l+1-2|\alpha_l|^2\right)\alpha_m\alpha^{\ast}_l+c.c.\right]\nonumber\\
&\le&\sum_{l,m,s}\mathcal{S}_{ml}\mathcal{S}_{ls}\left(2n_l+1-2|\alpha_l|^2\right)\alpha_m\alpha^{\ast}_s-\left[\sum_{l\ne m}\mathcal{S}_{ml}\mathcal{S}_{ll}\left(2n_l+1-2|\alpha_l|^2\right)\alpha_m\alpha^{\ast}_l+c.c.\right]\nonumber\\
&\le&\sum_{l,m,s}\mathcal{S}_{ml}\mathcal{S}_{ls}\left(2n_l+1-2|\alpha_l|^2\right)\alpha_m\alpha^{\ast}_s+2\left|\sum_{l\ne m}\mathcal{S}_{ml}\mathcal{S}_{ll}\left(2n_l+1-2|\alpha_l|^2\right)\alpha_m\alpha^{\ast}_l\right|,
\label{eq:(A. 10)}
\end{eqnarray}
where have dropped two positive-definite terms to get the first inequality and used the relation $a+a^{\ast}\le2|a|$ to get the second inequality. Defining $K_1= \sum_{l,m,s}\mathcal{S}_{ml}\mathcal{S}_{ls}\left(2n_l+1-2|\alpha_l|^2\right)\alpha_m\alpha^{\ast}_s$ and $K_2=2\left|\sum_{l\ne m}\mathcal{S}_{ml}\mathcal{S}_{ll}\left(2n_l+1-2|\alpha_l|^2\right)\alpha_m\alpha^{\ast}_l\right|$, then we have
\begin{eqnarray}
K_1&=&\sum_{l,m,s}\mathcal{S}_{ml}\mathcal{S}_{ls}\left(2n_l+1-2|\alpha_l|^2\right)\alpha_m\alpha^{\ast}_s\nonumber\\
&=&\braket{\bm{\alpha}\mathcal{S}\mathcal{N}^{\prime}\mathcal{S},\bm{\alpha}}\le\left|\braket{\bm{\alpha}\mathcal{S}\mathcal{N}^{\prime}\mathcal{S},\bm{\alpha}}\right|,
\label{eq:A11}
\end{eqnarray}
where $\bm{\alpha}=(\alpha_1,\alpha_2,\cdots,\alpha_{2d})$, $\braket{\bm{x},\bm{y}}\equiv \sum_j x_jy_j^{\ast}$ is the inner product of vectors $\bm{x}$ and $\bm{y}$, and $\mathcal{N}^{\prime}_{lm}=\delta_{lm}(2n_l+1-2|\alpha_l|^2)$. To obtain an upper bound on $K_1$, we use the inequalities 
\begin{eqnarray}
&&\left|\braket{\bm{x}O,\bm{y}}\right|\le||\bm{x}||\ ||O||\ ||\bm{y}||,\label{eq:inq1}\\
 &&||AB||\le  ||A|| \ ||B||,\label{eq:inq2}
\end{eqnarray}
where $ ||\cdot||$ is the norm of a vector or a matrix. For a matrix, $||O||$ is the largest eigenvalue of $O$ in the absolute value. Therefore, we have
\begin{eqnarray}
K_1\le2n_{\text{max}}w_{\max}^2\sum_{m}|\alpha_m|^2\le4dn_{\text{max}}w_{\max}^2\alpha_{\max}^2,\label{eq:A13}
\end{eqnarray}
where $n_{\text{max}}=\max\limits_l\left(n_l+1/2-|\alpha_l|^2\right)$, $w_{\max}\equiv \max_j|w_j|$ with $w_j$ the eigenvalues of $\mathcal{S}$, and $\alpha_{\max}=\max_l |\alpha_l|$. The last inequality from the above equation follows from $\sum_{m}|\alpha_m|^2\le 2d\alpha_{\max}$. For the second term in the last line in Eq. \eqref{eq:(A. 10)},

\begin{eqnarray}
K_2&=&2\left|\sum_{l,m}\mathcal{S}_{ml}\mathcal{S}_{ll}\left(2n_l+1-2|\alpha_l|^2\right)\alpha_m\alpha^{\ast}_l-\sum_{l}\mathcal{S}_{ll}^2\left(2n_l+1-2|\alpha_l|^2\right)|\alpha_l|^2\right|\nonumber\\
&\le&2\left|\sum_{l,m}\mathcal{S}_{ml}\mathcal{S}_{ll}\left(2n_l+1-2|\alpha_l|^2\right)\alpha_m\alpha^{\ast}_l\right|+2\left|\sum_{l}\mathcal{S}_{ll}^2\left(2n_l+1-2|\alpha_l|^2\right)|\alpha_l|^2\right|\nonumber\\\label{eq:A14}
&=&2\left|\braket{\bm{\alpha}\mathcal{S}\mathcal{S}^{\prime}\mathcal{N}^{\prime},\bm{\alpha}}\right|+2\left|\braket{\bm{\alpha}\mathcal{S}^{\prime2}\mathcal{N}^{\prime},\bm{\alpha}}\right|\nonumber\\
&\le&4n_{\text{max}}w_{\max}^2\sum_{m}|\alpha_m|^2+4n_{\text{max}}w_{\max}^2\sum_{m}|\alpha_m|^2\le16dn_{\text{max}}w_{\max}^2\alpha_{\max}^2,\end{eqnarray}
where $\mathcal{S}^{\prime}_{lm}=\delta_{lm}\mathcal{S}_{ll}$ and $||\mathcal{S}^{\prime}||\le w_{\max}$. 
Therefore, the fourth term in $\mathcal{F}_{\bm{w}}/4$ is
\begin{eqnarray}
F_4\le K_1+K_2\le 20dn_{\text{max}}w_{\max}^2\alpha_{\max}^2.
\end{eqnarray}

%where $\text{eigs}(\mathcal{S}^{\prime})_j=\sum_{l}\left|U_{jl}\right|^2w_l\le |w_{\max}|\sum_{l}\left|U_{jl}\right|^2=|w_{\max}|$. 

The fifth term in $\mathcal{F}_{\bm{w}}/4$ is expanded as
\begin{eqnarray}
F_5&=&\left[\sum_{l,m,s}\mathcal{S}_{ml}\mathcal{S}_{sl}\left(\xi_l^{\ast}-\alpha_l^{\ast 2}\right)\alpha_m\alpha_s-\sum_{l}\mathcal{S}_{ll}^2\left(\xi_l^{\ast}-\alpha_l^{\ast 2}\right)\alpha_l^2-\sum_{l\ne m}\mathcal{S}_{ml}^2\left(\xi_l^{\ast}-\alpha_l^{\ast 2}\right)\alpha_m^2-2\sum_{l\ne m}\mathcal{S}_{ml}\mathcal{S}_{ll}\left(\xi_l^{\ast}-\alpha_l^{\ast 2}\right)\alpha_m\alpha_l+c.c\right]\nonumber\\
&\le&2\left|\sum_{l,m,s}\mathcal{S}_{ml}\mathcal{S}_{sl}\left(\xi_l^{\ast}-\alpha_l^{\ast 2}\right)\alpha_m\alpha_s\right|+2\left|\sum_{l}\mathcal{S}_{ll}^2\left(\xi_l^{\ast}-\alpha_l^{\ast 2}\right)\alpha_l^2\right|+2\left|\sum_{l\ne m}\mathcal{S}_{ml}^2\left(\xi_l^{\ast}-\alpha_l^{\ast 2}\right)\alpha_m^2\right|+4\left|\sum_{l\ne m}\mathcal{S}_{ml}\mathcal{S}_{ll}\left(\xi_l^{\ast}-\alpha_l^{\ast 2}\right)\alpha_m\alpha_l\right|.\nonumber\\
\label{eq:A16}
\end{eqnarray}
We first define the terms on the right-hand side of the inequality in the above equation as $K_3, K_4,K_5$ and $K_6$, respectively. Then we have
\begin{eqnarray}
K_3&=&2\left|\sum_{l,m,s}\mathcal{S}_{ml}\mathcal{S}_{ls}\left(\xi_l^{\ast}-\alpha_l^{\ast 2}\right)\alpha_m\alpha_s\right|=2\left|\braket{\bm{\alpha}\mathcal{S}\mathcal{X}\mathcal{S},\bm{\alpha^{\ast}}}\right|\le 4d\xi_{\max}w_{\max}^2\alpha_{\max}^2,
\end{eqnarray}
where $\mathcal{X}_{lm}=\delta_{lm}\left(\xi_l^{\ast}-\alpha_l^{\ast 2}\right)$ and $\xi_{\max}\equiv\max\limits_j\left|\xi_j^{\ast}-\alpha_j^{\ast 2}\right|$. The inequality follows from Eq. \eqref{eq:inq1} and Eq. \eqref{eq:inq2}. Similarly, the second term in the above equation is 
\begin{eqnarray}
K_4&=&2\left|\sum_{l}\mathcal{S}_{ll}^2\left(\xi_l^{\ast}-\alpha_l^{\ast 2}\right)\alpha_l^2\right|=2\left|\braket{\bm{\alpha}\mathcal{S}^{\prime 2}\mathcal{X},\bm{\alpha^{\ast}}}\right|\le 4d\xi_{\max}w_{\max}^2\alpha_{\max}^2.
\end{eqnarray}
The third term is
\begin{eqnarray}
K_5&=&2\left|\sum_{l\ne m}\mathcal{S}_{ml}^2\left(\xi_l^{\ast}-\alpha_l^{\ast 2}\right)\alpha_m^2\right|\le2\xi_{\max}\alpha_{\max}^2\sum_lw_l^2,
\end{eqnarray}
where the inequality is obtained using the same reasoning as in Eq. \eqref{eq:S9}  . The fourth term on the right-hand side of the inequality in Eq. \eqref{eq:A16} is
\begin{eqnarray}
K_6&=&4\left|\sum_{l\ne m}\mathcal{S}_{ml}\mathcal{S}_{ll}\left(\xi_l^{\ast}-\alpha_l^{\ast 2}\right)\alpha_m\alpha_l\right|\le4\left|\sum_{l,m}\mathcal{S}_{ml}\mathcal{S}_{ll}\left(\xi_l^{\ast}-\alpha_l^{\ast 2}\right)\alpha_m\alpha_l\right|+2K_4\nonumber\\
&=&4\left|\braket{\bm{\alpha}\mathcal{S}\mathcal{S}^{\prime }\mathcal{X}^{\prime},\bm{\alpha^{\ast}}}\right|+2K_4\le16d\xi_{\max}w_{\max}^2\alpha_{\max}^2.
\end{eqnarray}
Therefore, the fifth term in $\mathcal{F}_{\bm{w}}/4$ is 
\begin{eqnarray}
F_5\le K_3+K_4+K_5+K_6\le 2\xi_{\max}\alpha_{\max}^2\sum_lw_l^2+24d\xi_{\max}w_{\max}^2\alpha_{\max}^2
\end{eqnarray}

Now we come to the last term in $\mathcal{F}_{\bm{w}}/4$, which is given by
\begin{eqnarray}
F_6&=&\sum_{l\ne s}\mathcal{S}_{ll}\mathcal{S}_{sl}\left(2\beta_l^{\ast}+\alpha_l^{\ast}-2n_l\alpha^{\ast}_l\right)\alpha_s+c.c.\nonumber\\
&\le&2\left|\sum_{l,s}\mathcal{S}_{ll}\mathcal{S}_{sl}\left(2\beta_l^{\ast}+\alpha_l^{\ast}-2n_l\alpha^{\ast}_l\right)\alpha_s-\sum_{l}\mathcal{S}_{ll}^2\left(2\beta_l^{\ast}+\alpha_l^{\ast}-2n_l\alpha^{\ast}_l\right)\alpha_l\right|\nonumber\\
&\le&2\left|\sum_{l,s}\mathcal{S}_{ll}\mathcal{S}_{sl}\left(2\beta_l^{\ast}+\alpha_l^{\ast}-2n_l\alpha^{\ast}_l\right)\alpha_s\right|+2\left|\sum_{l}\mathcal{S}_{ll}^2\left(2\beta_l^{\ast}+\alpha_l^{\ast}-2n_l\alpha^{\ast}_l\right)\alpha_l\right|\nonumber\\
&=&4\left|\braket{\bm{\alpha}\mathcal{S}\mathcal{S}^{\prime },\bm{\beta^{\prime}}}\right|+4\left|\braket{\bm{\alpha}\mathcal{S}^{\prime 2},\bm{\beta^{\prime}}}\right|,
\end{eqnarray} 
where $\bm{\beta^{\prime}}=\left(\beta_1+\alpha_1/2-n_1\alpha_1,\beta_2+\alpha_2/2-n_2\alpha_2,\cdots,\beta_{2d}+\alpha_{2d}/2-n_{2d}\alpha_{2d}\right)$. Using the inequalities in Eq. \eqref{eq:inq1} and Eq. \eqref{eq:inq2}, we have 
\begin{eqnarray}
\left|\braket{\bm{\alpha}\mathcal{S}\mathcal{S}^{\prime },\bm{\beta^{\prime}}}\right|&\le& w_{\max}^2\sqrt{\sum_l|\alpha_l|^2\sum_m|\beta_m+1/2-n_m|^2}\le2dw_{\max}^2\alpha_{\max}\beta_{\max},
\end{eqnarray}
where $\beta_{\max}=\max\limits_l\left|\beta_l+\alpha_l/2-n_l\alpha_l\right|$. Similarly, $\left|\braket{\bm{\alpha}\mathcal{S}^{\prime 2},\bm{\beta^{\prime}}}\right|\le2dw_{\max}^2\alpha_{\max}\beta_{\max}$. So
\begin{eqnarray}
F_6\le 16dw_{\max}^2\alpha_{\max}\beta_{\max}.
\end{eqnarray}
By collecting all the terms and replacing $w_{\max}=1/d$, we finally arrive at Eq. \eqref{eq:gBound} in the manuscript
\begin{eqnarray}
\mathcal{F}_{\bm{w}}\le  \frac{A}{d}+B|\bm{w}|^2,
\end{eqnarray}
where 
\begin{eqnarray}
A&=&16\alpha_{\max}\left(5n_{\max}\alpha_{\max}+6\xi_{\max}\alpha_{\max}+4\beta_{\max}\right),\\
B&=&4\left(v_{\text{max}}+M_{\max}+\Xi_{\max}+2\xi_{\max}\alpha_{\max}^2\right).
\end{eqnarray}
To get a simpler bound for $\mathcal{F}_{\bm{w}}$, we first consider the inequalities
\begin{align}
n_{\max}&\le \max_jn_j+1/2, \ & \xi_{\max}&\le\max_j|\xi_j|, \ & \beta_{\max}&\le \max_j|\beta_j|+\max_jn_j\max_k|\alpha_k|, \nonumber\\
v_{\max}&\le\max_jm_j, \ & M_{\max}&\le\left( \max_jn_j\right)^2+ \max_jn_j, \ &\Xi_{\max}&\le\left(\max_j|\xi_j|\right)^2,\label{eq:A31}
\end{align}
from the definitions of $n_{\max}, \xi_{\max}, \beta_{\max},v_{\max}, M_{\max}$ and $\Xi_{\max}$. Using Cauchy-Schwarz inequality, we have
\begin{align}
|\alpha_j|&\le \sqrt{n_j}, &n_j& \le \sqrt{ m_j}, &|\xi_j|&\le \sqrt{m_j-n_j}\le \sqrt{m_j} , &|\beta_j|&\le \sqrt{m_jn_j}.\label{eq:A32}
\end{align}
Using these inequalities, we obtain an upper bound for $A$ as
\begin{align}
A\le304\max_jm_j+40\max_jn_j,
\end{align}
and an upper bound for $B$ as
\begin{align}
B&\le20\max_jm_j+4\max_jn_j.
\end{align}
Note that $m_j=\braket{a_j^{\dagger}a_j^{\dagger}a_ja_j}+\braket{a_j^{\dagger}a_j}\ge n_j$, hence we have $A\le 344\max_jm_j$ and $B\le 24\max_jm_j$. Defining $\braket{n^2}_{\max}=\max_jm_j$ and for well-distributed weights $|\bm{w}|\le 1/d$, we obtain
\begin{eqnarray}
\mathcal{F}_{\bm{w}}<\frac{C^2}{d}\braket{n^2}_{\max},
\end{eqnarray}
where the smallest integer satisfies the above inequality is $C=20$. Plugging the above result into Eq. \eqref{eq:q_bound}, we arrive at Eq. \eqref{eq:ngt}.

\end{widetext}

\end{document}